\documentclass[preprint2]{aastex62}
\usepackage[T1]{fontenc}
\usepackage{graphicx, url, times}
\usepackage{booktabs, threeparttable}
\usepackage{multirow}
\usepackage{amsmath}

\graphicspath{{pics/}}

\newcommand{\papertitle}{Estimation of the Galaxy Quenching Rate in the Illustris Simulation}

\newcommand{\hkpc}{{\ifmmode{h^{-1}{\rm kpc}}\else{$h^{-1}$kpc}\fi}}

\newcommand{\Fig}[1]{Figure~\ref{#1}}
\newcommand{\Eqn}[1]{Equation~\ref{#1}}
\newcommand{\Tbl}[1]{Table~\ref{#1}}
\newcommand{\Sec}[1]{Section~\ref{#1}}
\newcommand{\Ssec}[1]{Subsection~\ref{#1}}

\def\Msun{M_{\odot}}

\newcommand{\affa}{School of Physics and Astronomy, Sun Yat-sen University, 519082, Zhuhai, China}

\received{March 20, 2020}
\revised{October 24, 2020}
\accepted{November 18, 2020 }
\submitjournal{ApJ}
\shorttitle{Quenching Rate in Illustris}
\shortauthors{Yang et al.}

\newlength{\figwidth}
\newlength{\resplot}
\hypersetup{ pdfauthor={Yang Wang}, pdftitle={\papertitle},
pdfsubject={pdfsub}, urlcolor=blue, }

\begin{document}

\title[]{\papertitle}
\correspondingauthor{Yang Wang}
\email{wangyang23@mail.sysu.edu.cn}

\author[0000-0002-1512-5653]{Yang Wang}
\affil{\affa}

\author{Xuan Liu}
\affil{\affa}

\author{Weishan Zhu}
\affil{\affa}

\author{Lin Tang}
\affil{\affa}

\author{Weipeng Lin}
\affil{\affa}

\label{firstpage}
\begin{abstract}

	Quenching is a key topic in exploring the formation and evolution of galaxies.
	In this work, we study the quenching rate, i.e., the variation in the fraction of quenched galaxies per unit time, of the Illustris-1 simulation.
	By building the quenched fraction function $f(m,\rho, t)$ of each snapshot in the simulation, we derive an accurate form of quenching rate as $\Re_q=df(m,\rho,t)/dt$.
	According to the analytic expression of the quenching rate $\Re_q$, we split it into four components: mass quenching, environmental quenching, intrinsic mass quenching and intrinsic environmental quenching.
	The precise value and evolutions can be given via the formula of $\Re_q$.

	With this method, we analyze the Illustris-1 simulation.
	We find that quenched galaxies concentrate around $M_*\simeq10^{11}h^{-1}M_\odot$ and  $\delta+1\simeq10^{3.5}$ at earlier times, and that the quenching galaxy population slowly shifts  to lower stellar mass and lower overdensity regions with time.
	We also find that mass quenching dominates the quenching process in this simulation, in agreement with some previous analytical models.
	Intrinsic quenching is the second most important component.
	Environmental quenching is very weak, because it is possible that the pre- or postprocessing of environments disguises environmental quenching as intrinsic quenching.

	We find that our method roughly predict the actual quenching rate. It could well predict the actual amount of galaxies quenched by intrinsic  quenching. However, it overestimates the amount of mass quenching galaxies and underestimates the amount of environmental quenching. We suggest that the reason is the nonlinearity of the environmental overdensity change and mass growth of the galaxy.

\end{abstract}

\keywords{
	methods: numerical  -- galaxies: evolution
}

\section{Introduction}
\label{sec:intro}
According to current observational data, galaxies population shows a clear bimodality on the color-color (or equivalently color-magnitude, color-mass, SFR-mass) map \citep{Balogh2004,Baldry2004b,Baldry2006,Cassata2008,Wetzel2012}.
This bimodal distribution divides galaxies into two categories: star-forming galaxies with strong star formation activities that are younger and bluer and  exhibit late-type morphologies, and quiescent galaxies with little star formation activities that are older and redder and exhibit early-type morphologies \citep{Blanton2003,Kauffmann2003a,Noeske2007,Wel2014}.
Observational studies suggest that a mechanism for shutting down star formation, i.e., quenching, is required to realize the evolution of luminosity function (or equivalently, stellar mass function) of red sequence and blue cloud galaxies \citep{Arnouts2007,Faber2007}.
In addition to their color, these two galaxy populations differ in many other galaxy properties, such as metallicity, age and morphology.
Therefore, researchers agree that a relation exists between quenching and galaxy properties \citep{Kauffmann2004,Brinchmann2004,Muzzin2012,Muzzin2013,Pallero2019}.
Overall, galaxy quenching plays an important role in galaxy formation and evolution.

Galaxy quenching is considered to be driven by various physical processes.
Usually, a galaxy is quenched if its gas is exhausted or prevented from cooling .
Stellar winds or super nova explosions could blow away gas\citep{Larson1974,Dekel1986,DallaVecchia2008}.
AGN feedback from a central supermassive black hole could heat or even remove the gas in a galaxy\citep{Croton2006,Fabian2012,Fang2013e,Cicone2014,Bremer2018}.
Moreover, gas could be stripped by interactions between galaxies and their environments.
For example, the hot gas will be completely removed when its host galaxy is accreted into a larger system, i.e., becomes a satellite galaxy, which is called `starvation' or `strangulation'\citep{Larson1980}.
Moreover, ram pressure  stripping can sweep cold gas out of a galaxy when it travels through a high density region with high speed\citep{Gunn1972}, and harassments caused by high-speed encounters between two galaxies can deplete cold gas rapidly by accelerating star formation\citep{Moore1996, Mihos2004}.

The most common method for exploring the influence of these physics is investigating the quenching efficiency in relation to the stellar mass, environments, and time.
It is commonly believed that the internal physics should depend on some intrinsic quantities of a galaxy such as stellar mass, bulge mass or black hole mass, while the external physics should depend on environmental parameters such as local overdensity, halo mass and centric distance from the cluster center \citep{Silk1998, Peng2010,Contini2020}.
Therefore, a relation between quenching and galaxy stellar/bulge/black-hole mass, and a relation between quenching and the environmental parameters are expected.
When the physics are evolved with time , we expect an evolved  or a static quenching-mass-environment relation.
Most works have reached the consensus that galaxies with higher stellar mass or in denser regions have higher quenching efficiency\citep{Peng2010,Balogh2016,Darvish2016,PintosCastro2019,Contini2019}.
The main issue  now is to quantify the quenching with respect to different underlying mechanisms. Therefore, the separation of different components in the quenching process is urgently needed.

On the other hand, the evolution of quenching efficiency is of increasing concern because it provide valid information on what roles different
physical mechanisms play in different epochs.
In earlier works quenching efficiency is thought to be unchanged with time \citep{Peng2010}.
In contrast, some recent works claimed an evolved quenching efficiency \citep{PintosCastro2019,Kawinwanichakij2017,Quadri2012,Contini2020}.

Motivated by the debates above, we seek to  link the quenching history more closely to the physics behind quenching.
In previous works, the widely used parameter ``quenching efficiency'' in fact reflects a cumulative subsequence of all physics in the past.
The instantaneous physical activities are more likely to correlate with the speed of quenching, i.e., quenching rate.
Exploring the quenching rate could help us quantify the influence of different quenching mechanisms at different times.
However, it is somewhat difficult to build up the full history of galaxies purely with observational data.
In recent years, simulations such as Illustris\citep{Vogelsberger2014}, EAGLE\citep{Schaye2015}, Illustris-TNG\citep{Pillepich2018}, and SIMBA\citep{Dave2019}, have well reproduced the population of quenched galaxies to some extent, in agreement with observations.
These simulations provide us with an excellent frame for building up the histories of galaxies and exploring associated quenching events.

In this work, we represent a method for deriving an analytical formula of the quenching rate from the data of galaxy populations of different redshifts.
We expect this analytical method to predict a quenching rate reflecting the actual galaxy history. Therefore, we apply it to the data of the Illustris-1 simulation and test its validation.

We will introduce the data and methods we used in \Sec{sec:data}.
In \Sec{sec:quench0} we briefly summarize the relation between quenched fraction, stellar mass and overdensity at redshift $0$.
Then, in the next section, we state the method we used to obtain a fitting model of the evolving quenching rate across cosmic time and the main conclusions we could derive from this fitting model. This is the main section presenting our results.
In \Sec{sec:compare}, we further discuss  how our model matches the actual histories of galaxies, how to link our model with physics mechanisms and which  factors possibly affect the accuracy of this model.
Finally a summary is presented in \Sec{sec:con}.

\section[]{Data and Methods}
\label{sec:data}
We apply our analysis to the Illustris-1 data.
The Illustris-1 simulation is a cosmological hydrodynamic simulation with a comoving volume of $(106.5{\rm Mpc})^3$.
Its cosmological parameters are consistent with the WMAP9 data release \citep{Hinshaw2013}, assuming that $\Omega_{\Lambda}=0.7274$, $\Omega_{m}=0.2726$, $\Omega_b=0.0456$, $\sigma_8=0.809$ and $H_0=70.4\rm{km\ s^{-1} Mpc^{-1}}$.
The simulation contains $1820^3$ dark matter particles , $1820^3$ hydrodynamic cells, and $1820^3$ Monte Carlo tracer particles. The softening length is $1420\ \rm{pc}$ for dark matter and $710\ \rm{pc}$ for baryon particles and gas cells. The mass resolution of dark matter particles is $6.26\times10^6M_{\odot}$ and the mean mass resolution of baryons is $1.26\times10^6M_{\odot}$.

The simulation evolves the initial condition from redshift $127$ to $0$ with $136$ output snapshots ($2$ of which are broken),
including the treatment of gravitation, hydrodynamics and astrophysical processes such as gas cooling and photoionization, star formation, ISM modeling, stellar evolution, stellar feedback and AGN feedback.
Finally, it resolves $4366546$ substructures at redshift $0$.
For details of the simulation, readers can refer to \cite{Vogelsberger2014a,Vogelsberger2014,Genel2014,Sijacki2015}.

The Illustris project provides complete (sub)halo catalogue and merger trees on their website \url{https://www.illustris-projet.org}.
The (sub)halos are identified by the \textsc{Subfind} algorithm \citep{Springel2001,Dolag2009}. Following previous works \citep[e.g.][]{Genel2014,Sparre2015}, the stellar components within one \textsc{Subfind} halo are recognized as one galaxy.
In this work, our analysis is restricted to galaxies with stellar masses larger than $10^9h^{-1}M_{\odot}$.
In the Illustris-1 simulation, galaxies above this mass limitation contain more than $1000$ stellar particles and are therefore well resolved to avoid most numerical uncertainties.
There are $11356$ galaxies above the mass limitation at $z=0$, $5524$ at $z=0.576$ and $2894$ at $z=1.2$.
The merger trees of \textsc{Subfind} halos are created using \textsc{SubLink} \citep{Rodriguez-Gomez2015}.
We take advantage of the \textsc{SubLink} merger tree to track the history of individual galaxies.

In this work we focus on the galaxy quenching process which is closely related to the stellar mass, star formation rate (SFR) and environmental overdensity. Thus, it is necessary to clarify how we derived these properties from the simulation data. We briefly list our methods in the following:

\begin{figure*}[htbp]
	\centering
	\includegraphics[width=0.4\linewidth]{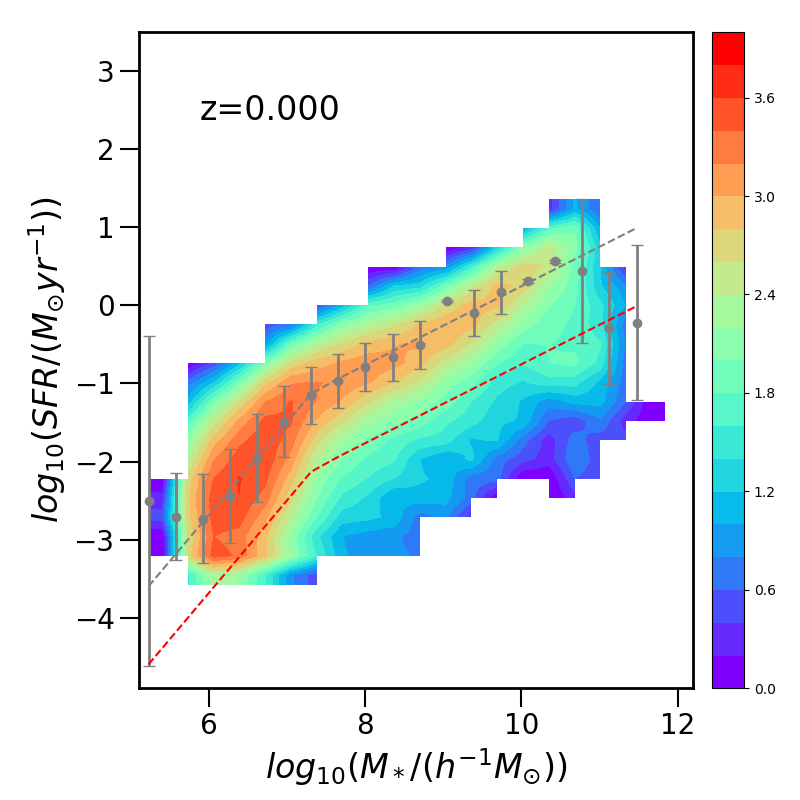}
	\includegraphics[width=0.4\linewidth]{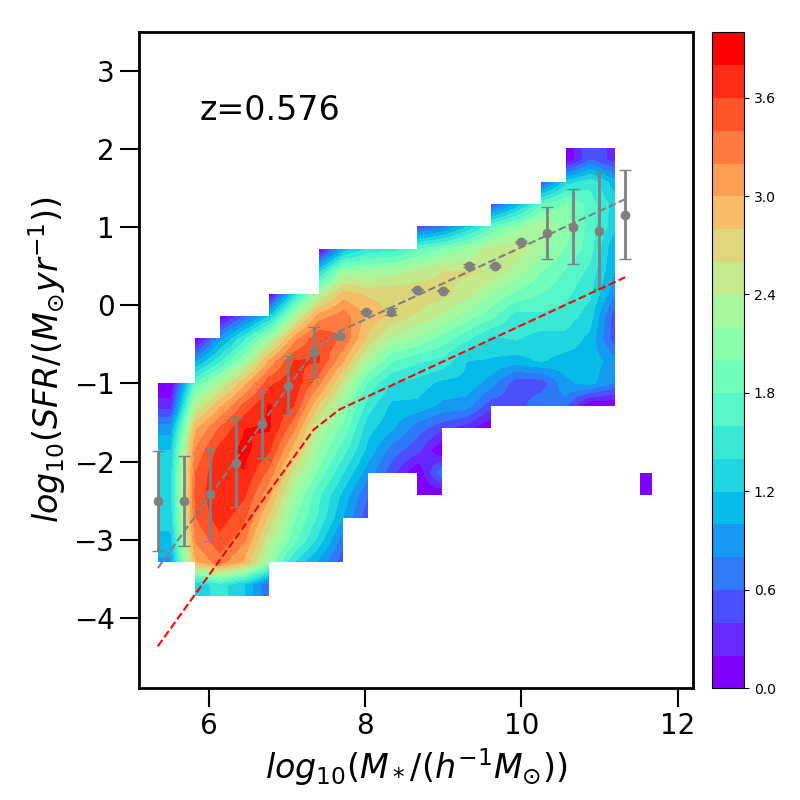}
	\includegraphics[width=0.4\linewidth]{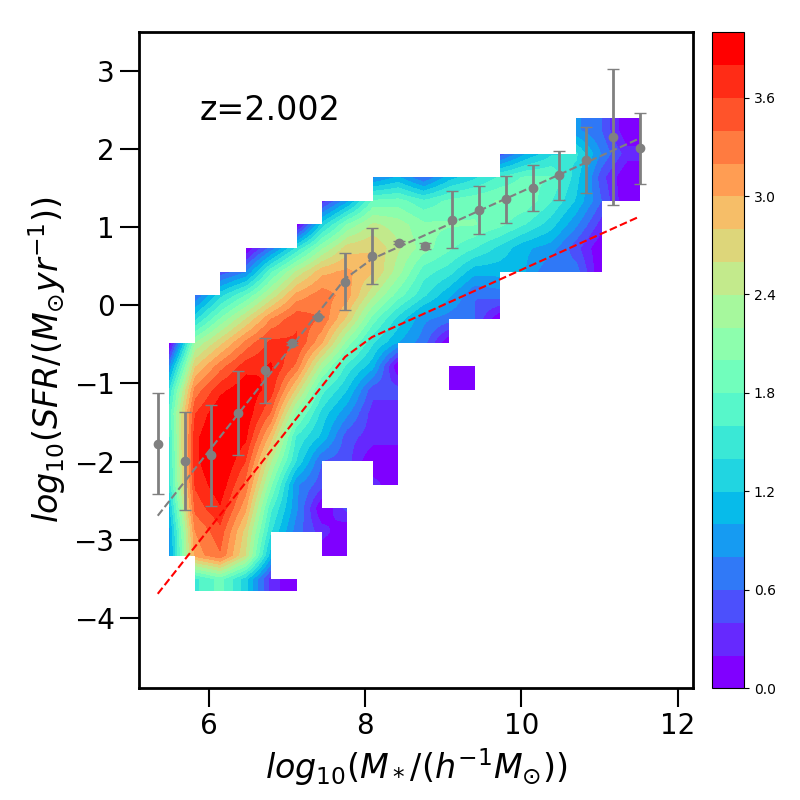}
	\includegraphics[width=0.4\linewidth]{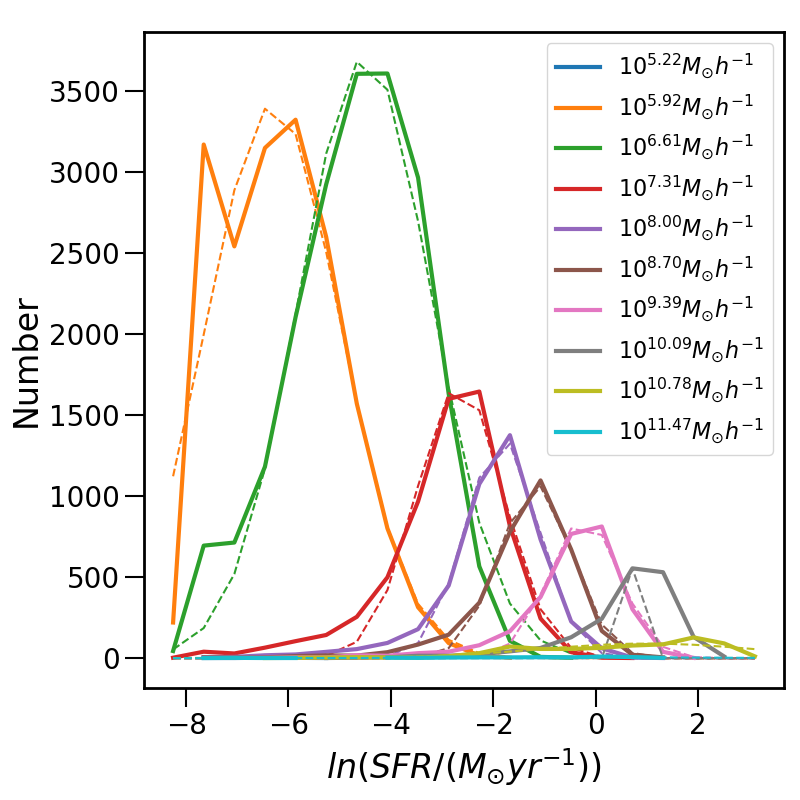}
	\caption{The contour maps of the $SFR$-$M_*$ distribution at different redshifts in Illustris-1. The grey points are the peaks of the SFR distribution in the corresponding stellar mass bin. The grey lines are fitting lines of the peaks of the SFR distribution, defined as the main sequence line. The red lines are quenching lines. Galaxies below the quenching lines are defined as quenched galaxies. Note that a large number of galaxies with a SFR of $0$ are not shown in these plots. The bottom-right subplot shows the distribution of the SFR in each mass bin at $z=0$. The dashed lines in the bottom-right subplot represent Gaussian fittings to the SFR distribution. }
	\label{sfrmap}
\end{figure*}

\begin{description}
	\item[Galactic Stellar Mass] The Illustris project provides galactic stellar mass with several definitions. To be comparable with observations, we use the data block `SubhaloStellarPhotometricsMassInRad'.
	      Specifically, the database gives the radius at which the surface brightness profile drops below the limit of $20.7\ \rm{mag\ arcsec^{-2}}$ in the K band. The data in the block `SubhaloStellarPhotometricsMasInRad' is the stellar mass within this radius.
	\item[Star Formation Rate] We use the sum of the star formation rates of all gas cells in a subhalo as the SFR of the galaxy this subhalo hosts. These data are denoted as `SubhaloSFR' in the Illustris subhalos catalogue.
	\item[Environmental Overdensity] The environmental overdensity at a specific point is defined as $\delta (\vec r)=(\Sigma(\vec r)-\bar{\Sigma})/\bar{\Sigma}$,
	      where $\Sigma$ is the galaxy number density and $\bar{\Sigma}$ is the mean galaxy number density of the whole universe.
	      Since it is very difficult to measure the actual matter density of a certain point in the universe, researchers usually use the galaxy number density instead of the true matter density.
	      There are varies methods for computing the galaxy number density at one point. Readers could refer to \cite{Muldrew2012} for a detailed description of them.
	      Here, we use a method mimic the approach taken in \citep{Peng2010}.
	      For a specific point, we find its $5th$ nearest bright neighbor galaxies within an aperture of $\pm 1000 kms^{-1}$ along the $z$-axis; here, `bright' means that their $r$ band magnitudes are brighter than $-19.5$.
	      With the projected distance between the central point and its $5th$ nearest neighbor $R_5$, we obtain the field density as $\Sigma =\frac{5}{20 {\rm Mpc}h^{-1}\times\pi R_5^2} $.
	      On the other hand, the mean density is calculated as $\bar{\Sigma}=\frac{N_{gal}}{(75{\rm Mpc}h^{-1})^3}$, in which $N_{gal}$ is the total number of galaxies brighter than $M_r=-19.5$.

	\item[How to define `quench'] The definition of `quench' varies in different works.
	      Many observers draw a division line between the red peak and blue peak on the color-mass diagram to distinguish between star-forming and quenched galaxies \citep[e.g.][]{Blanton2005b,Peng2010,Muzzin2013}.
	      The first piece of work of the Illustris project \cite{Vogelsberger2014} adopted the same criteria as \cite{Blanton2005b}.
	      Some works \citep{Steinborn2015,Wang2018b,RodriguezMontero2019,DeLucia2019, Franx2008}  used a simple definition that quenched galaxies have a specific star formation rate ($sSFR=SFR/M_*$) smaller than $0.2/t_H(z)$, where $t_H(z)$ is the age of the universe at redshift $z$.
	      Another work \cite{Bluck2016}, which compared the quenched galaxies in the SDSS survey, the Illustris simulation and the L-Galaxy simulation,
	      defined quenched galaxies as SFR-$M_*$ scatters  below  the main SFR-$M_*$ sequence minus $1\ dex$. \cite{Donnari2019, Donnari2020} compared the criteria of the main sequence cut,  as in \cite{Bluck2016}, and the UVJ cut on the UVJ diagram. They found that these two methods result in very similar populations of quenched and star-forming galaxy populations in the Illustris TNG100 simulation.
	      In this work we use a method similar to that of \cite{Bluck2016} .
	      \Fig{sfrmap} shows how we determine the quenching line.
	      First, we find the main sequence line on the $SFR$-$M_*$ plane for each snapshot.
	      It is defined as a fitting line through the peaks of galaxy SFR distributions.
	      (grey lines in \Fig{sfrmap}).
	      The $ln SFR$ of galaxies in each stellar mass bin obeys a Gaussian distribution, as the right bottom subplot in \Fig{sfrmap} shows.
	      By fitting the formula,  the peaks of SFR distribution in each mass bin are obtained as the grey dots in the plots of \Fig{sfrmap}.
	      The error bars of grey dots show the range of $1$ standard deviation of the fitted Gaussian formula.
	      Then, we use a linear function to fit the position of the peaks.
	      We find that the main sequence lines at difference snapshots in Illustris-1 have  two parts that follow a linear function for all redshifts.
	      The quenching lines are defined by shifting the main sequence lines $1$ $dex$ down, shown as the red lines in \Fig{sfrmap}.
	      All galaxies below the quenching lines are attributed to quenched galaxies.
	      In \Fig{sfrmap}, we show the main sequence line and quenching line for all galaxies in the snapshots, but keep in mind that in the following content, our analysis focuses only on galaxies more massive than $10^9h^{-1}M_\odot$.
\end{description}

\section{Quenching Fraction at Different Redshifts}
\label{sec:quench0}

\begin{figure*}[htbp]
	\centering
	\includegraphics[width=0.45\linewidth]{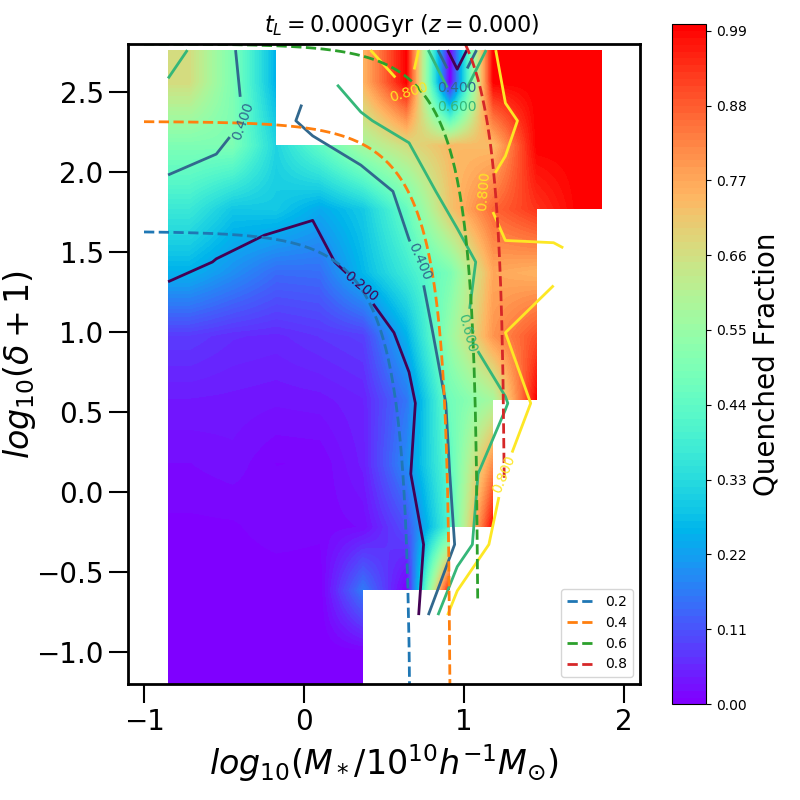}
	\includegraphics[width=0.45\linewidth]{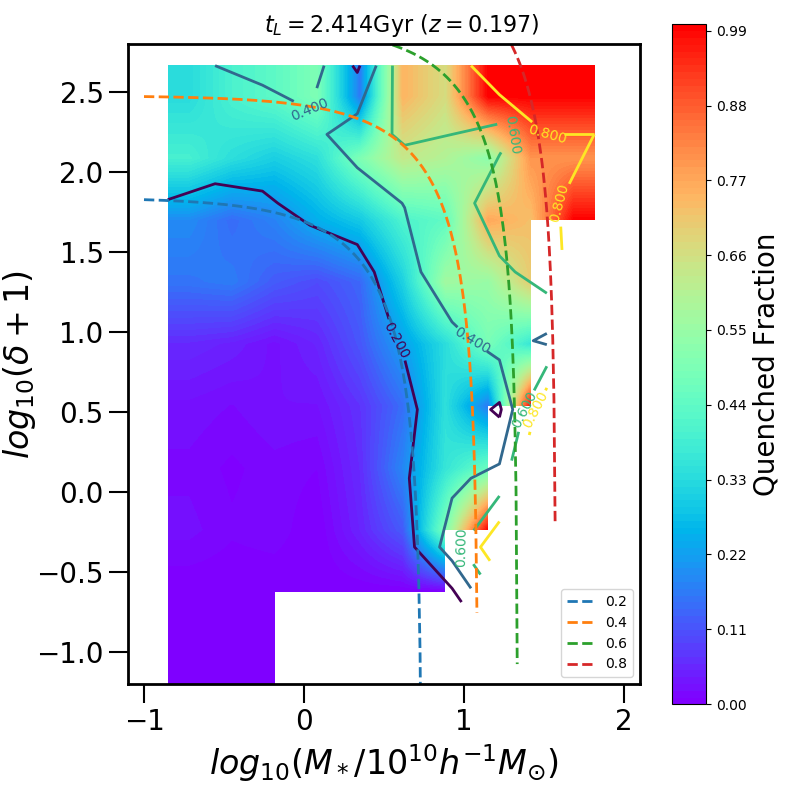}
	\includegraphics[width=0.45\linewidth]{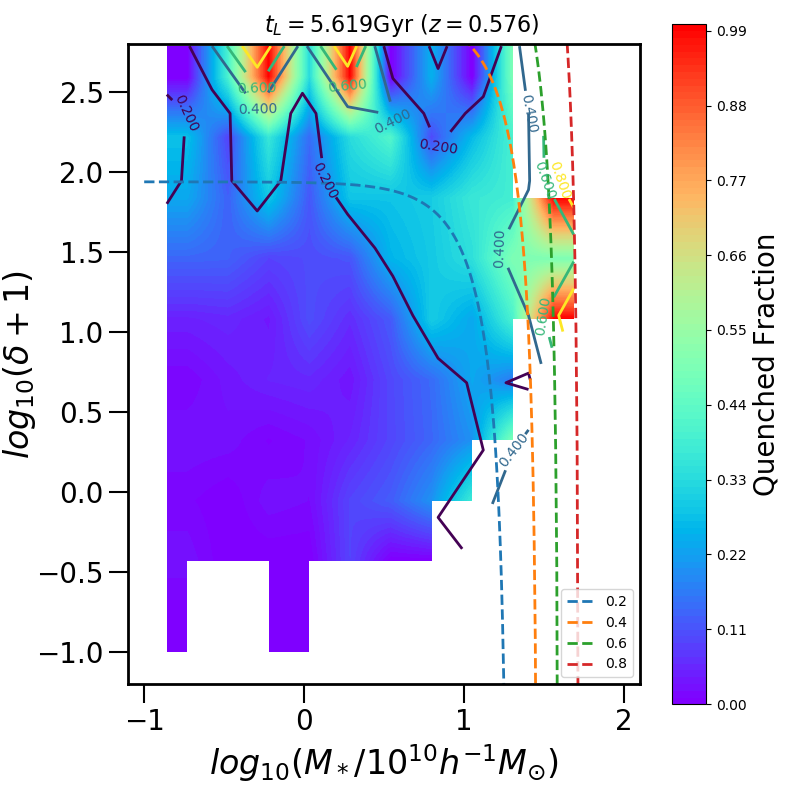}
	\includegraphics[width=0.45\linewidth]{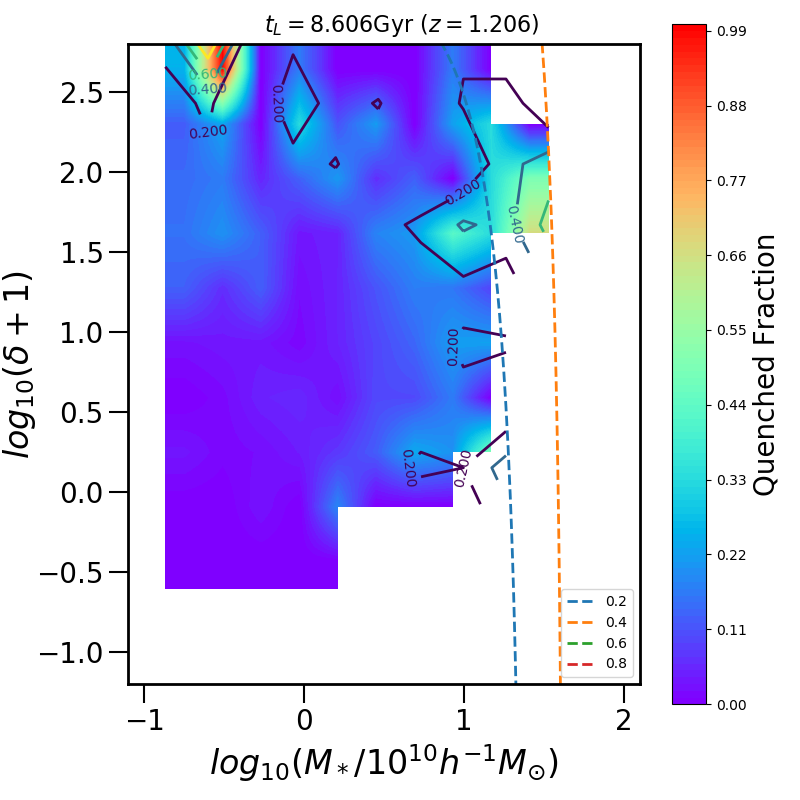}
	\caption{The quenched fraction in Illustris-1 as a function of the stellar mass and environmental overdensity at redshift $z=0$, $z=0.197$,$z=0.576$ and $z=1.206$.
		Lines with numbers are contours of the actual SFR distribution.
		The colored dashed lines indicate fitted contours of different quenched fractions given by our fitting function.
		The corresponding quenched fraction values are shown in the legend.
	}
	\label{FigQF}
\end{figure*}
Initially, we try to reproduce the quenched fraction as function of stellar mass and environmental overdensity.
As \Fig{FigQF} shows, the quenched fraction increases with stellar mass and environmental overdensity of the galaxy.
The same trend can also be seen in Figure 6 of \cite{Peng2010} and Figure 16 of \cite{Vogelsberger2014}.
According to \cite{Peng2010}, we use the following function to fit the quenched fraction $f_q$ of galaxies in Illustris-1:
\begin{equation}
	f_{q}(\rho,m) =1-e^{-(\frac{\rho}{\rho_c})^{\alpha_\rho}-(\frac{m}{m_c})^{\alpha_m}}
	\label{EquFr}
\end{equation}
In the following context, if not specified, $m$ refers to the stellar mass of galaxies in units of $10^{10}h^{-1}M_\odot$,  and $\rho$ refers to $1+\delta$.

\begin{figure}[htbp]
	\centering
	\includegraphics[width=1\linewidth]{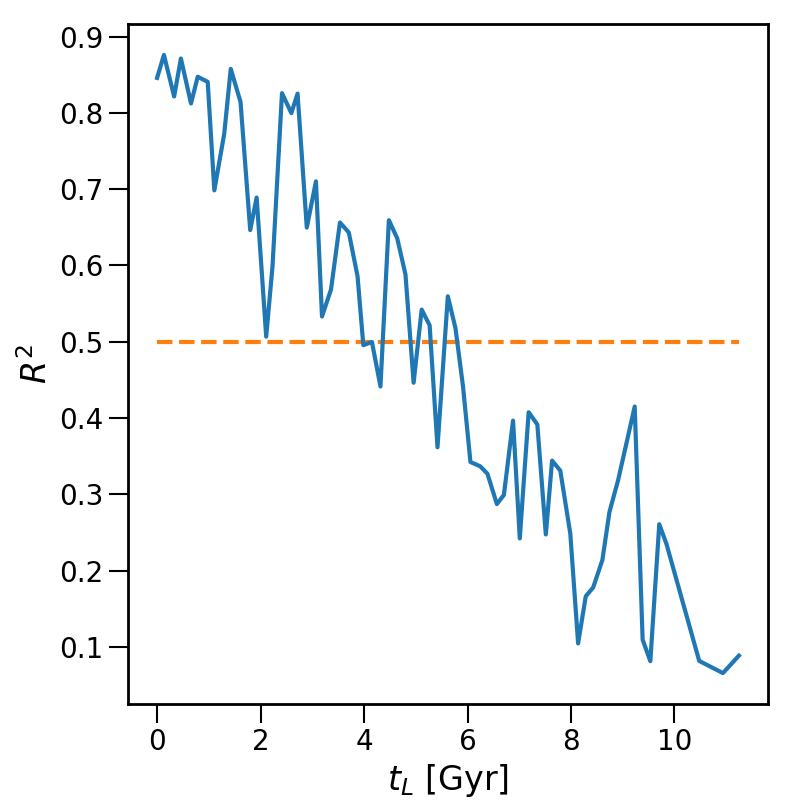}
	\caption{The goodness of fit, $R^2$, to the quenched fraction of galaxies in  each snapshot of Illustris-1.
		$x$ axis shows the looking back time $t_L$ of each snapshot.
		Snapshots that failed to be fitted were excluded.
	}
	\label{FigR2}
\end{figure}

The fitting lines are shown as dashed lines in \Fig{FigQF}.
As we can speculate from \Fig{FigQF}, \Eqn{EquFr} fits well at low redshifts.
When  $z>0.5$ ($t_L\gtrsim5\ Gyr$), active galaxies almost dominate the whole galaxy population, therefore blurring the changing trends of the quenched fraction.
To qualify the  fittings, we calculate the $R^2$ parameter of each snapshot.
$R^2$ is a parameter qualifying the goodness of fit which  ranges from $0$ to $1$.
The closer it is to $1$, the better the fitting.
Usually, a fitting with $R^2$ larger than $0.5$ is regarded as a good fitting.
As \Fig{FigR2} shows, the $R^2$ parameters are above $0.5$ for snapshots with a looking back time  $t_L<5\ Gyr$ (which corresponds to redshift $z\lesssim0.5)$.
For snapshots above $z=0.5$, the number of galaxies above $10^9 M_{\odot}h^{-1} $ is small, and the number of quenched galaxies is even smaller.
Therefore the statistics on the galaxy population have very large uncertainty.
On the other hand, we could claim that the distribution of quenched galaxies could meet the form of \Eqn{EquFr} well when $z<0.5$.

\begin{table*}[htbp]
	\centering
	\caption{The fitting parameters of the function of quenched fractions of the Illustris-1 simulation at $z=0, 0.197, 0.361, 0.576$ and $1.206$ and the fitting parameter for SDSS DR7 $0.02<z<0.085$ from \cite{Peng2010}. Note that  the $log(m_c)$ in \cite{Peng2010} is $10.56$, with $m_c$ in unit of $\Msun$, while in our fitting formula $m_c$ is in units of $10^{10}h^{-1}\Msun$. We have converted the values from \cite{Peng2010} to make them consistent  with our work, assuming $h=0.7$. }
	\label{TabFr}
	\begin{tabular}{c|cccc}
		\toprule
		sample                & $log(\rho_c)$            & $\alpha_\rho$  & $log(m_c)$               & $\alpha_m$     \\
		\hline
		SDSS DR7 0.02<z<0.085 & $1.84\pm0.01$            & $0.60\pm0.01$  & $0.71\pm0.01$            & $0.80\pm0.01$  \\
		Illustris-1 z=0       & $2.87^{+0.118}_{-0.163}$ & $0.52\pm0.098$ & $1.12^{+0.037}_{-0.040}$ & $1.46\pm0.238$ \\
		Illustris-1 z=0.197   & $2.99^{+0.129}_{-0.184}$ & $0.56\pm0.119$ & $1.37^{+0.046}_{-0.051}$ & $1.02\pm0.145$ \\
		Illustris-1 z=0.380   & $3.50^{+0.267}_{-0.816}$ & $0.38\pm0.116$ & $1.80^{+0.114}_{-0.156}$ & $1.02\pm0.326$ \\
		Illustris-1 z=0.576   & $3.88^{+0.291}_{-1.340}$ & $0.34\pm0.080$ & $1.61^{+0.049}_{-0.055}$ & $1.92\pm0.472$ \\
		Illustris-1 z=1.206   & $6.64^{+0.370}_{-1.643}$ & $0.19\pm0.089$ & $1.83^{+0.171}_{-0.287}$ & $1.43\pm0.655$ \\
		\bottomrule
	\end{tabular}
\end{table*}

We list the values of the fitting parameters of several snapshots in \Tbl{TabFr}. To compare with previous works, the parameters from \cite{Peng2010} are listed in the first line of the same table.
We find that although the quenched fraction distribution of Illustris-1 is in the same term as that in \cite{Peng2010}, the specific parameters are quite different.
The galaxies in Illustris-1 require a higher characteristic stellar mass and environmental overdensity than those in \cite{Peng2010} .
As \Tbl{TabFr} shows, the turning points $m_c$ and $\rho_c$ are $0.2\ dex$  and $1.68\ dex$ larger than those in \cite{Peng2010}.
This result indicates that in the Illustris-1 universe, galaxies tend to be quenched with higher stellar mass or environmental overdensity.
On the other hand, the slope of the quenching efficiency become  steeper for mass quenching but slightly shallower for environmental quenching.
This means that the intensity of the quenching process tends to increase with stellar mass more significantly.

These results are also reported in \cite{Vogelsberger2014}.
They claimed that  the shape of red fraction contours of the Illustris data could be well fitted by
shifting the contour of \cite{Peng2010}  $+0.1\ dex$ in mass and $+0.7\ dex$ in overdensity.
In \cite{Vogelsberger2014}, the contour line for the red fraction of $0.4$ in Illustris-1
coincides with the contour line for the red fraction of $0.9$ in \cite{Peng2010} plus a $0.1\ dex$ shift in mass and  a $0.7\ dex$ shift in overdensity.
This means that the red fraction in Illustris-1 and \cite{Peng2010} share the same term of fitting function but with different parameters.
Basically, in our analysis, the distribution of the quenched fraction in mass-overdensity space is quite similar to that of the red fraction in \cite{Vogelsberger2014}.
Considering that we have only a difference in the threshold defining quenched/red galaxies, it is not surprising to obtain these results.
From another point of view, defining quenched galaxies by the sSFR or by the color does not seem to make much of a difference.

\section{Quenching Rate Derived from Multiple Snapshots}
\label{sec:quenchz}

In this work our main purpose is to explore how the quenched fraction $f_q$ changes with time.
Our method is to fit $f_q$ at multiple redshifts in the Illustris-1 simulation and then to determine the quenching rate as $\Re_q=df_q/dt$.
In our work, $t$ is chosen to be the universal time in units of $Gyr$; therefore,
all quenching rates mentioned hereafter are in units of $Gyr^{-1}$.
\cite{Peng2010} made a similar attempt but stopped half-way because the change in quenched fraction across time is so small.
Building the history of the quenched fraction in observations requires combining data from different surveys, which brings system differences larger than the changes in the quenched fraction.
Thus, the authors decided to look into the evolution of the mass function of star-forming galaxies instead of the quenching efficiency.
However, with full maps of the galaxy history in one simulation, for Illustris-1, it is possible to explore the issues of galaxy quenching from this viewpoint.

\subsection{Analytic Formula of the Quenching Rate}

\begin{figure}[htbp]
	\centering
	\includegraphics[width=1\linewidth]{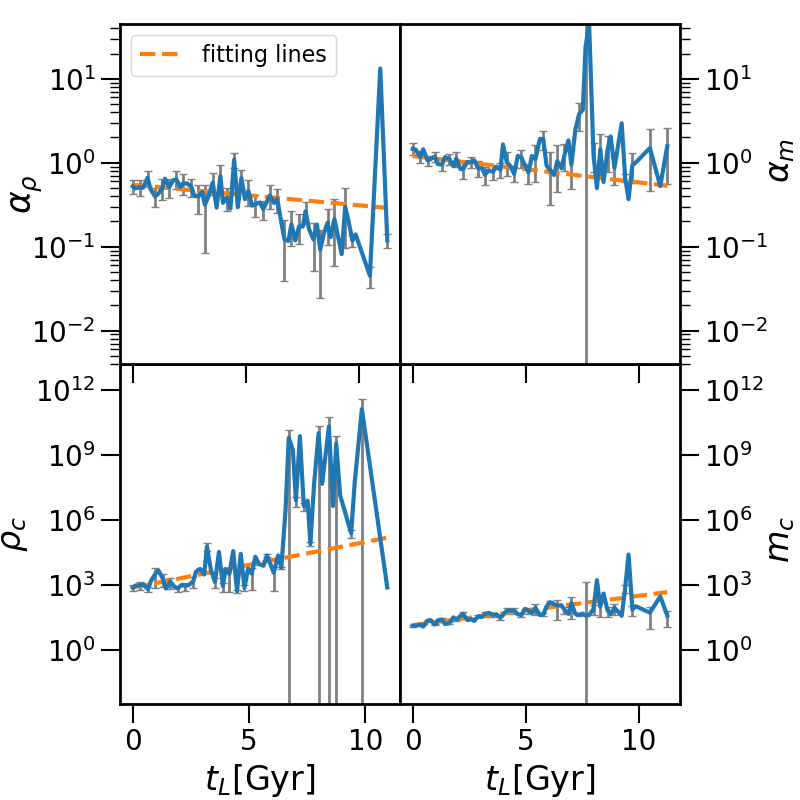}
	\caption{Four parameters of the fitted quenched fraction function ($m_c$,$\alpha_{m}$,$\rho_c$, $\alpha_{\rho}$) as functions of the looking back time $t_L$ (blue lines). The error bars show the uncertainty in estimating these parameters. The fitting lines of their evolving curves are given as orange dashed lines.}
	\label{FigParaQF}
\end{figure}

In \Sec{sec:quench0}, we found that the quenched fraction could be well fitted by \Eqn{EquFr} from $t_L=0$ to $t_L=5$.
For higher looking back times, we can still use the fitting term of \Eqn{EquFr}, although large uncertainty is introduced.
No obvious clue suggests that the quenched fraction at earlier times exhibits a different distribution.
We plot four fitting parameters at different $t_L$ for all successfully fitted snapshots in \Fig{FigParaQF}.

\Fig{FigParaQF} shows very obvious trends of four parameters of the quenched fraction function.
Basically, $m_c$, and $\rho_c$ increase with redshifts, while $\alpha_m$ and $\alpha_\rho$ decrease .
These trends are the same as those of \cite{Peng2010},
though the parameters in the Illustris-1 simulation change much more significantly .
This result implies that the galaxy quenching process in the simulation may be too intensive compared with the observations.
However it is also possible that observational uncertainty conceals the evolution trends of galaxy quenching.
A discussion on this topic would require additional investigations to be carried out; hence, in this work, we do not discuss this topic any further.

With the assumption that quenched fraction functions have the same form at different times, we assume that there is a uniform fitting function $f_q(m, \rho, t)$ for all redshifts.
Then, the velocity of quenching, i.e., the quenching rate, could be calculated by $\Re_q=df_q/dt$.
In principle, the quenching rate could tell us what fraction of galaxies are quenched per unit time.
The variable $t$ in the function $f_q$ is introduced by the evolution of $m$, $\rho$, and the parameters $m_c, \alpha_m, \rho_c and \alpha_\rho$.
Therefore, the quenching rate could be expanded in the following way:
\begin{equation}
	\label{EquRq}
	\begin{aligned}
		\Re_q=\frac{df_q}{dt}= & \frac{\partial f_q}{\partial m}\frac{\partial m}{\partial t} +         \\
		                       & \frac{\partial f_q}{\partial \rho}\frac{\partial \rho}{\partial t} +   \\
		                       & \frac{\partial f_q}{\partial m_c}\frac{\partial m_c}{\partial t}
		+\frac{\partial f_q}{\partial \alpha_m}\frac{\partial \alpha_m}{\partial t} +                   \\
		                       & \frac{\partial f_q}{\partial \rho_c}\frac{\partial \rho_c}{\partial t}
		+\frac{\partial f_q}{\partial \alpha_\rho}\frac{\partial \alpha_\rho}{\partial t}               \\
		=                      & \lambda_m+\lambda_\rho+\Re_{q,im}+\Re_{q,i\rho}
	\end{aligned}
\end{equation}
Note that the variable $t$ in \Eqn{EquRq} is the universal time.
To make things more convenient we also use the looking back time $t_L$ in this work,
which is in the opposite direction relative to the universal time.
The variable $t_L$ will appear frequently in the equations in following context.
Note that we have to apply $\Re_q=df/dt=-df/dt_L$ when combining the equation with $t_L$ to \Eqn{EquRq}.

\begin{figure}[htbp]
	\centering
	\includegraphics[width=1\linewidth]{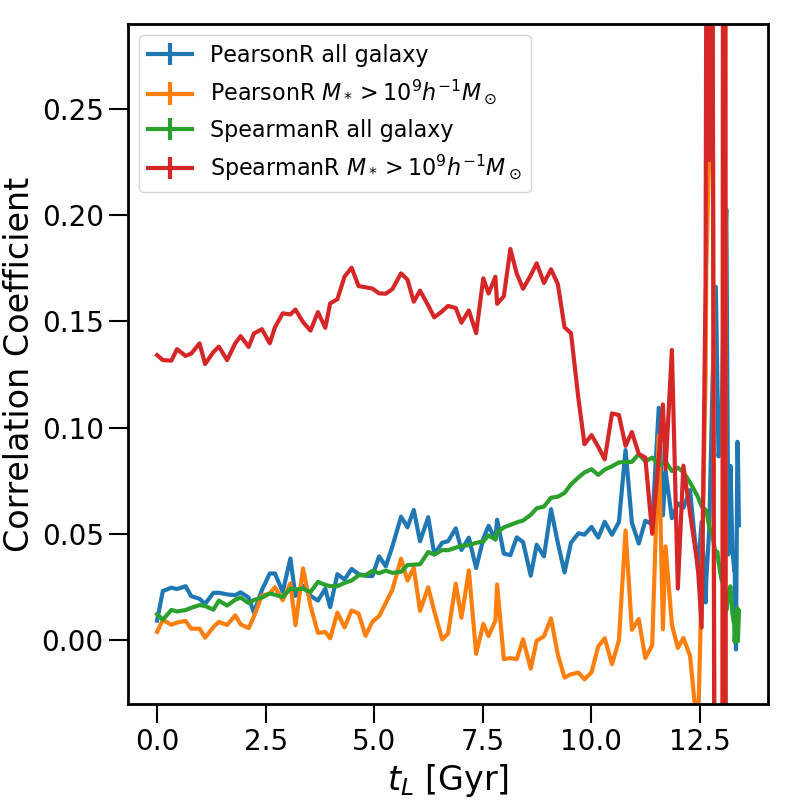}
	\caption{The correlation coefficient between the stellar mass and environmental overdensity of galaxies in Illustris-1 at different times.
	The blue and orange lines show the Pearson correlation coefficient.
	The green and red lines show the Spearman correlation coefficient.
	The orange and red lines show the correlation coefficient only for galaxies more massive than $10^{9}h^{-1}M_\odot$, while the remaining two show the results for all galaxies. }
	\label{FigCC}
\end{figure}

In \Eqn{EquRq}, there is an underlying condition that  $m$, $\rho$, $m_c$, $\alpha_m$, $\rho_c$ and $\alpha_\rho$ are only functions of $t$, not conceivably of each other of them.
The parameters $m_c$, $\alpha_m$, $\rho_c$ and $\alpha_\rho$ should not be functions of $m$ or $rho$.
They are fitting parameters for a specific range of stellar mass and environmental overdensity .
Hence, these four parameters should be independent of any of the others by default.
For $m$ and $\rho$, it is common to assume that an $m-\rho$(stellar mass - environmental overdensity) relation exists.
Intuitively, larger galaxies tend to reside in higher-density regions.
However, this trend only sets a boundary of the $m-\rho$ distribution, rather than a close $m-\rho$ relation.
We use the correlation coefficient to check the relation between $m$ and $\rho$, as \Fig{FigCC} shows.
The correlation coefficient of $m$ and $\rho$ remains very low.
The correlation coefficient ranges from $0$ to $1$.
The higher the values is, the stronger the relation between two variables.
We check two kinds of coefficient: the Pearson correlation coefficient and the Spearman correlation coefficient.
The former one can qualify the strength of a linear relation, while the latter can represent non-linear relation.
We also use different galaxy populations to check the $m-\rho$ relation.
From \Fig{FigCC}, we can see that only galaxies above $10^9h^{-1}M_\odot$(red line) have slightly higher correlation coefficients (around $0.16$), while the other three lines are close to $0$.
\Fig{FigCC} shows an almost null correlation between the galaxy stellar mass and environmental overdensity.
There might be a very small nonlinear relation of $m-\rho$ for galaxies more massive than $10^9h^{-1}M_\odot$.
However, this relation is too weak to show its significance.
Therefore, $m$ and $\rho$ are independent on each other, and any item containing $\partial{m}/\partial{\rho}$ or $\partial{\rho}/\partial{m}$ can be ignored from \Eqn{EquRq}.

Apparently, the expanded equation can be separated into four parts:
\begin{enumerate}
	\item The $\partial f_q/\partial m\cdot\partial m/\partial t$ part stands for the changes in quenched fraction caused by the growth in the galaxy stellar mass, with the other five parameters held constant , i.e., the transformation of quenched galaxies from one mass bin to another or a quenching process with mass growth (including continuous  growth and merger growth), denoted as the mass quenching rate $\lambda_m$ hereafter.
	\item The $\partial f_q/\partial \rho \cdot \partial \rho/\partial t$  part stands for the changes in quenched fraction caused by migration to different environments, i.e., the transformation of quenched galaxies from one overdensity bin to another or the quenching process  accompanied by environmental changes such as satellite quenching by gas stripping,  denoted as the environmental quenching rate $\lambda_\rho$ hereafter.
	\item The part containing $m_c$ and $\alpha_m$ stands for a change in the quenched fraction due to the time evolution of $m_c$ and $\alpha_m$ when the stellar mass of the galaxy is fixed. This part may relate to some  intrinsic galaxy properties that correlate with the stellar mass or sensitive to are some characteristic mass, i.e. AGN feedback or stellar winds. We denote it as the intrinsic mass quenching rate $\Re_{q,im}$.
	\item The part containing $\rho_c$ and $\alpha_\rho$ stands for a change in the quenched fraction when the environmental overdensity is fixed. This part relates to the environmental origin physics, not taking into account the environmental changes, i.e. merger rate, or relates to the delayed influence from the environment. We denote it as the intrinsic environmental quenching rate $\Re_{q,i\rho}$.
\end{enumerate}
We remind readers that the terms we used here are not fully consistent with previous definitions.
In previous works, ``mass quenching'' usually refers to quenching from internal physics, such as feedbacks, while ``environmental quenching'' usually refers to external physics driven quenching, such as merger quenching or satellite quenching.
Our definition is based on the mathematical format of items in the analytical formula of the quenching rate.
Therefore it can not distinguish between internal and  external sources.
The physical criterion of our definition is whether quenching occurs together with stellar mass changes or environmental overdensity changes.
THese criteria could be strictly constrained in math.
However, the difference is not very large between our definition and previous ones.
Basically, in this work, ``mass quenching'' $+$ ``intrinsic mass quenching'' is equivalent to mass quenching(internal quenching) plus some contribution from merger quenching in definitions from previous works.
``Environmental quenching'' $+$ ``intrinsic environmental quenching'' is equivalent to most previous environmental quenching(external quenching) concepts.
We consider merger quenching to contribute to both the mass and environmental parts because merger will result in mass growth, while the merger rate is affected by the environment.

In the following, we discuss the intrinsic quenching rate, mass quenching rate and environmental quenching rate separately in three subsections.

\subsection{Intrinsic Quenching Rate}
Combining \Eqn{EquFr} and \Eqn{EquRq}, we can express the formulas of intrinsic quenching rate as:
\begin{equation}
	\label{EquRim}
	\begin{split}
		\Re_{q,im} = &e^{-(\frac{\rho}{\rho_c})^{\alpha_\rho}-(\frac{m}{m_c})^{\alpha_m}}\times \hfill \\
		&(\frac{m}{m_c})^{\alpha_m}(ln(\frac{m}{m_c})\frac{\partial \alpha_m}{\partial t}-\frac{\alpha_m}{m_c}\frac{\partial m_c}{\partial t}) \hfill
	\end{split}
\end{equation}
\begin{equation}
	\label{EquRip}
	\begin{split}
		\Re_{q,i\rho} = &e^{-(\frac{\rho}{\rho_c})^{\alpha_\rho}-(\frac{m}{m_c})^{\alpha_m}}\times \hfill \\
		&(\frac{\rho}{\rho_c})^{\alpha_\rho}(ln(\frac{\rho}{\rho_c})\frac{\partial \alpha_\rho}{\partial t}-\frac{\alpha_\rho}{\rho_c}\frac{\partial \rho_c}{\partial t}) \hfill
	\end{split}
\end{equation}

\begin{figure*}[htbp]
	\centering
	\includegraphics[width=1\linewidth]{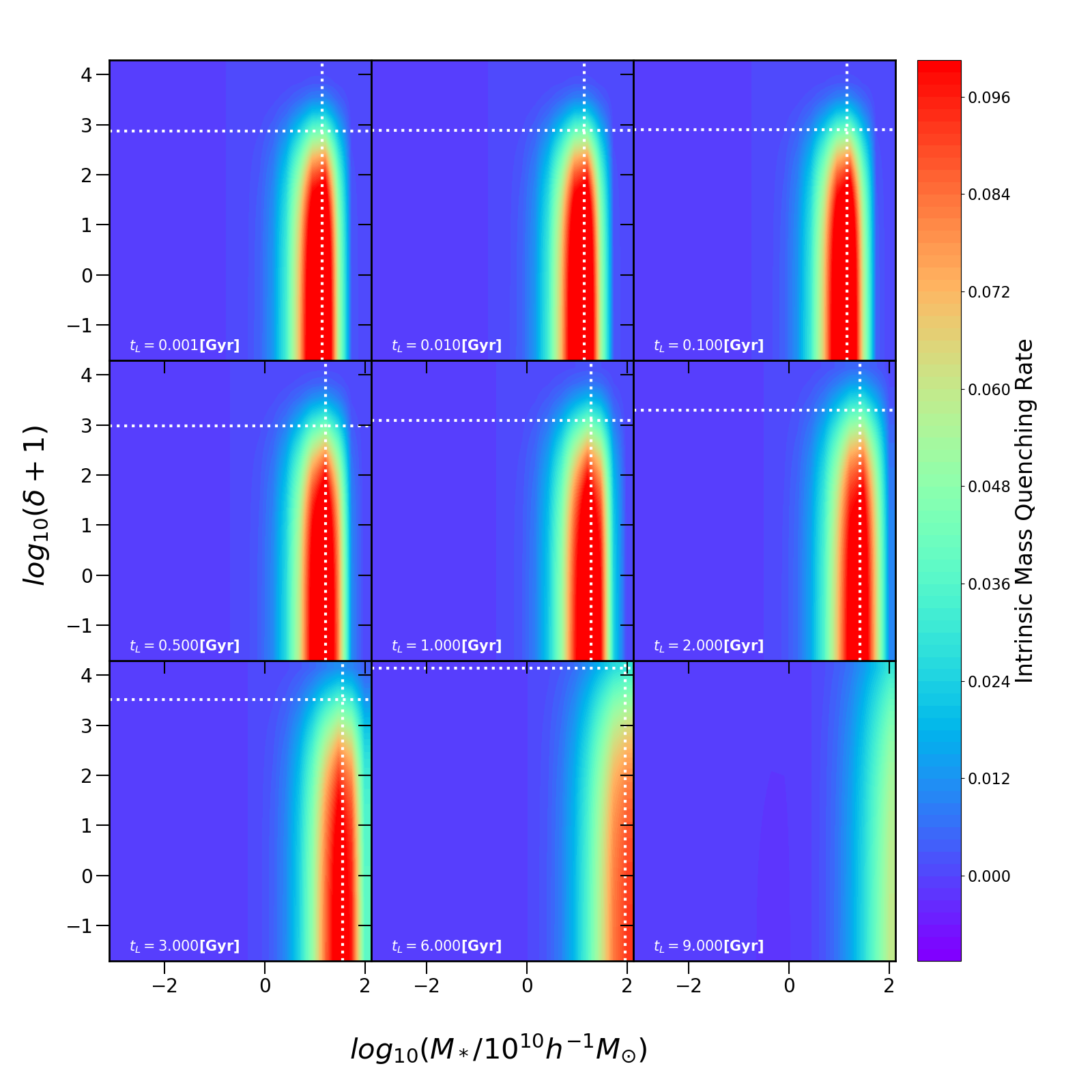}
	\caption{The  intrinsic mass quenching rate $\Re_{q,im}$ as a function of the stellar mass and environmental overdensity of galaxies at different times.
		The value of $\Re_{q,im,}(m,\rho,t_L)$ is represented by the color.
		The white dotted lines show $m_c(t)$ and $\rho_c(t)$ at that time.
	}
	\label{FigRqIm}
\end{figure*}

\begin{figure*}[htbp]
	\centering
	\includegraphics[width=1\linewidth]{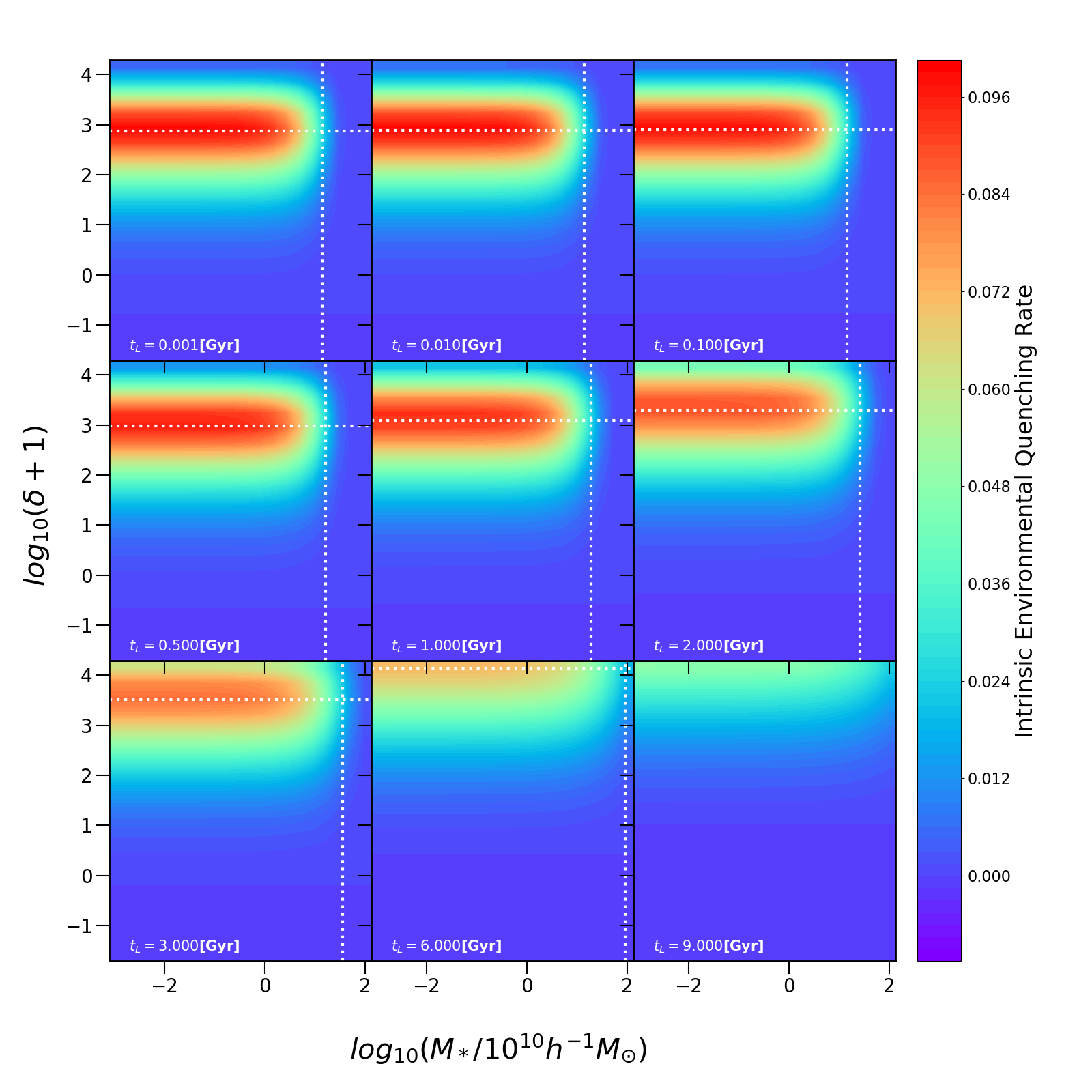}
	\caption{The  intrinsic environmental quenching rate $\Re_{q,i\rho}$ as a function of the stellar mass and environmental overdensity of galaxies at different times. The configuration is the same as in \Fig{FigRqIm}.
		The white dotted lines show $m_c(t)$ and $\rho_c(t)$ at that time.}
	\label{FigRqId}
\end{figure*}

As \Eqn{EquRim} and \Eqn{EquRip} show, the problem of evaluating the intrinsic quenching rate turns into  obtaining the time dependent function  $m_c(t)$, $\alpha_m(t)$, $\rho_c(t)$ and $\alpha_\rho(t)$.
To achieve these four functions, we plot the fitting curves (orange dashed lines) in \Fig{FigParaQF}.
Because the fitted quenched fraction function has large uncertainty when $t_L>5Gyr$, we give more weight to the points at $t_L\le5Gyr$.
We found that all parameters could be fitted with the term $log_{10}(p) = at+c$:
\begin{equation}
	\label{EquFitFr}
	\begin{split}
		&	log_{10}(m_c) = 0.136t_L+1.14\\
		&	log_{10}(\alpha_m)= -0.0315t_L+0.0801\\
		&	log_{10}(\rho_c)=0.211t_L+2.87\\
		&	log_{10}(\alpha_\rho)=-0.0245t_L-0.263\\
	\end{split}
\end{equation}

After inserting \Eqn{EquFitFr} into \Eqn{EquRim} and \Eqn{EquRip}, we obtain the final form of the functions of $\Re_{q,im}(m,\rho,t)$ and $\Re_{q,i\rho}(m,\rho,t)$:

\begin{equation}
	\begin{aligned}
		\Re_{q,im} & = (1-f_q(m,\rho,t_L)){10^{-0.0315t_L}}                              \\
		           & \times  (7.24\times 10^{-0.136t_L-2}m)^{1.20\times 10^{-0.0315t_L}} \\
		           & \times  (0.0872ln(10^{-0.136t_L}m)+0.148)                           \\
	\end{aligned}
	\label{EquRimFull}
\end{equation}
\begin{equation}
	\begin{aligned}
		\Re_{q,i\rho} & = (1-f_q(m,\rho,t_L)){10^{-0.0245t_L}}                                         \\
		              & \times  (1.35\times 10^{-0.211t_L^{1.87}-3}\rho)^{0.546\times 10^{-0.0245t_L}} \\
		              & \times  (0.0308ln(10^{-0.211t}\rho)+0.0617)                                    \\
	\end{aligned}
	\label{EquRipFull}
\end{equation}

\Eqn{EquRimFull} and \Eqn{EquRipFull} are  too complex to provide a simple view of the quenching rate; thus, we plot the distribution of the intrinsic mass quenching rate and intrinsic environmental quenching rate of $9$ snapshots in  \Fig{FigRqIm} and \Fig{FigRqId}.

As \Fig{FigRqIm} shows, the intrinsic mass quenching rate is dependent only on the galactic stellar mass, except that the null correlation occurs at the high overdensity end.
A $0$ quenching rate at the high overdensity end is naturally reasonable because the galaxies there are always $100\%$ quenched, resulting in no increment of the quenched fraction.
The $\Re_{q,im}$ has one peak following $m_c(t)$.
The position of the peak is at approximately $M_*=10^{12.0}h^{-1}\Msun$ at an early time (approximately $6$ Gyr ago), then slowly shifts to the position of  $M_*=10^{11.4}h^{-1}\Msun$ at present.
This implies that only galaxies within a very narrow mass range undergo the intrinsic mass quenching procedure,
and quenching galaxies are less massive than those in at earlier times.
The peak value of $\Re_{q,im}$ is approximately $0.09\ Gyr^{-1}$ at $t_L=6Gyr$ and grows slowly to $0.13\ Gyr^{-1}$ in present, implying an accelerating quenching.
In in other regions except the peak, $\Re_{q,im}$ is very small ($< 0.001$).

On the other hand, as \Fig{FigRqId} shows, the intrinsic environmental quenching rate $\Re_{q, i\rho}$ is dependent only on the environmental overdensity, except that the null correlation exhibited at the high stellar mass end.
The reason for the $0$ quenching rate at the high mass end is the same as that for $\Re_{q,im}$.
The $\Re_{q,i\rho}$ peaks at approximately $\delta=10^{4.13}$ at an earlier time (approximately $6$ Gyr ago), then  shifts to the position of  $\delta = 10^{2.87}$.
Similar to $\Re_{q,im}$, the peak position of $\Re_{q,i\rho}$ follows $\rho_c(t)$.
It implies that the intrinsic environmental quenching procedure also takes place in a narrow range of galaxies,
and intrinsic environmental quenching galaxies reside in regions less dense regions than those at earlier times.
Its peak values grow from $0.081\ Gyr^{-1}$ to $0.12\ Gyr^{-1}$ in the period from $t_L=9Gy$ to $t_L=0.001Gyr$.
At the same time, its peak position moves from $log(\delta+1)\approx4$ to $log(\delta+1)\approx3$.
It is also worth noting that the value of $\Re_{q,i\rho}$ is consistently lower than $\Re_{q,im}$ by a factor of $0.3\sim0.5$ in the regions outside the peaks.
This result indicates  that the environment has a relatively smaller influence on intrinsic galaxy quenching.

\subsection{ Mass Quenching Rate}
\label{SsecMQR}
The mass quenching rate $\lambda_m$ simply contains two parts: $\partial f_q/\partial m$ and $\partial m/\partial t$.
The former one can be easily evaluated from \Eqn{EquFr}, resulting in a expanded form of $\lambda_m$ as follows:
\begin{equation}
	\label{EquRM}
	\begin{aligned}
		\lambda_m = & \frac{\partial f_q}{\partial m}\frac{\partial m}{\partial t}                         \\
		=           & (1-f_q)\frac{\alpha_m}{m_c}(\frac{m}{m_c})^{\alpha_m-1}\frac{\partial m}{\partial t} \\
		=           & (1-f_q(m,\rho,t_L))\frac{1.20\times10^{-0.0315t_L}}{m}                               \\
		            & \times(7.24\times10^{-0.136t_L-2}m)^{1.20\times10^{-0.0315t_L}}                      \\
		            & \times\frac{\partial m}{\partial t}
	\end{aligned}
\end{equation}
Apparently, the difficulty in obtaining the exact value of \Eqn{EquRM} is determining the average mass growth rate $\partial m/\partial t$.

\begin{figure}[htbp]
	\centering
	\includegraphics[width=1\linewidth]{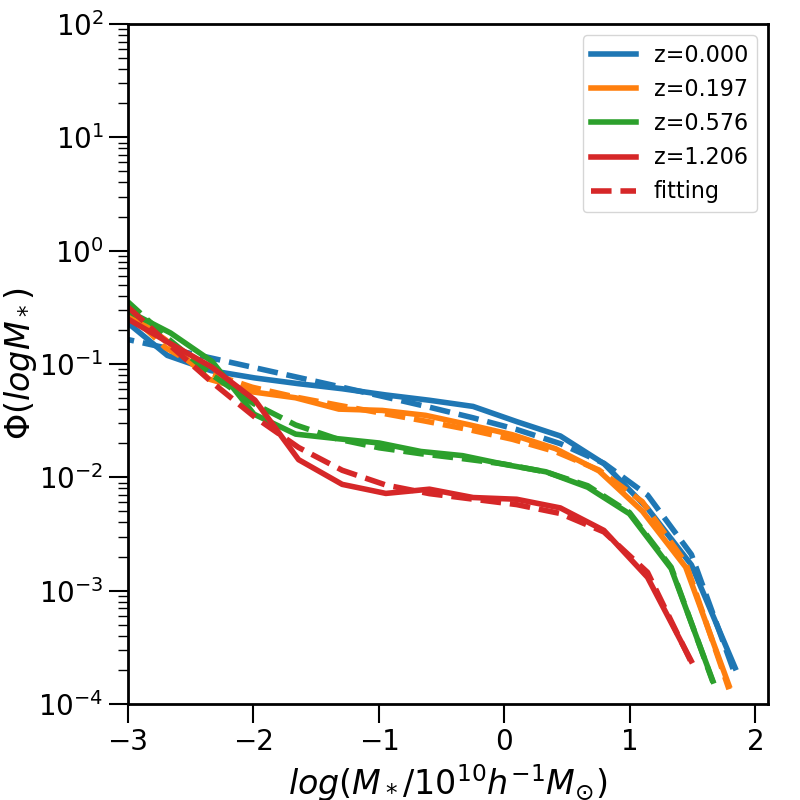}
	\caption{Mass function of galaxies in Illustris-1 at different redshifts. Dashed lines with the same color are fitting curves at the same redshifts.}
	\label{FigMF}
\end{figure}

\begin{figure}[htbp]
	\centering
	\includegraphics[width=1\linewidth]{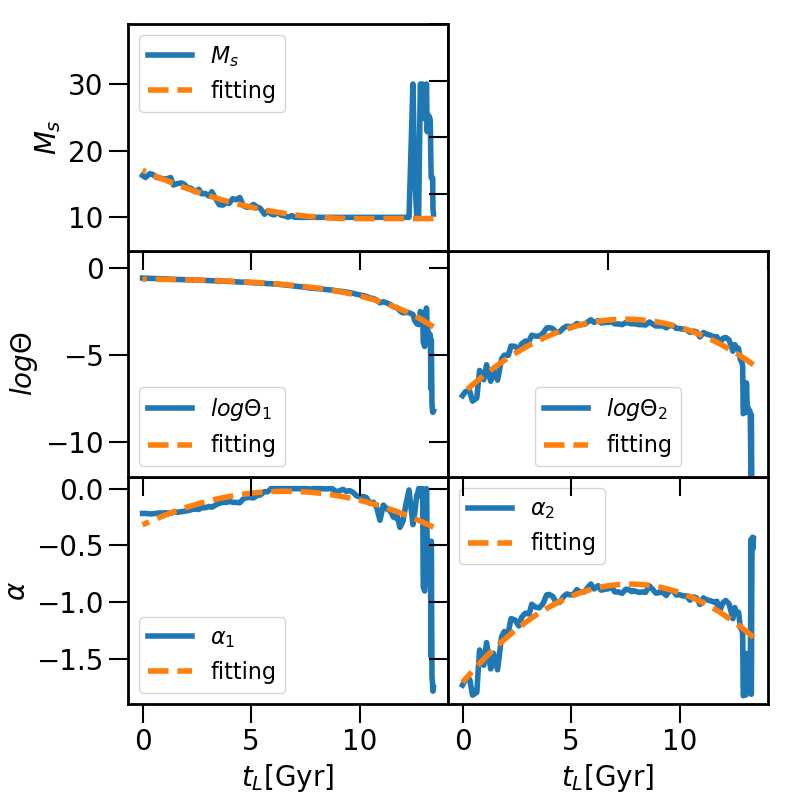}
	\caption{The parameters of the galaxy mass function as functions of the looking back time $t_L$. The fitting lines of their evolving curves are presented as orange dashed lines. }
	\label{FigParaMF}
\end{figure}

\begin{figure*}[htbp]
	\centering
	\includegraphics[width=1\linewidth]{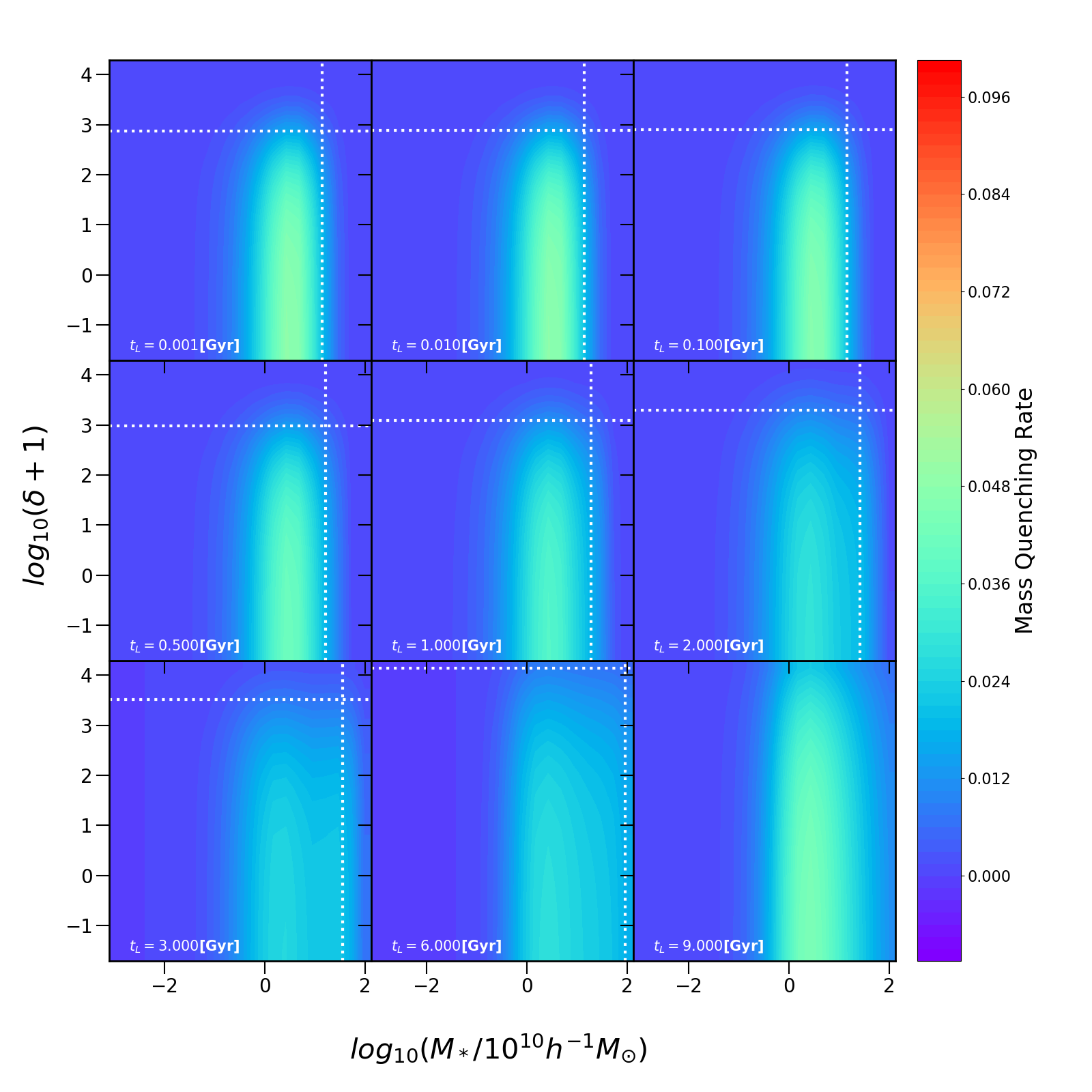}
	\caption{The mass quenching rate $\lambda_m$ as a function of the stellar mass and environmental overdensity of galaxies at different times. The configuration is the same as in \Fig{FigRqIm} except that the color bar has a larger range.
		The white dotted lines show  $m_c(t)$ and $\rho_c(t)$ at that time.
	}
	\label{FigRqM}
\end{figure*}

The main concept of our method is to derive this average mass growth rate via the galaxy stellar mass function.
The galaxy stellar mass function changes over time because galaxies continue to grow.
Conversely, if we know the changes in the stellar mass function, we can create a model of average mass growth of galaxies.

We find that the stellar mass function at different times can be well fitted by two components Schechter function:
\begin{equation}
	\begin{aligned}
		 &  & \Phi_m(m,t)= & \frac{\Theta_1(t)}{M_s(t)}(\frac{m}{M_s(t)})^{\alpha_1(t)}e^{-\frac{m}{M_s(t)}}+ \\
		 &  &              & \frac{\Theta_2(t)}{M_s(t)}(\frac{m}{M_s(t)})^{\alpha_2(t)}e^{-\frac{m}{M_s(t)}}
	\end{aligned}
	\label{EquMF}
\end{equation}
We create fittings of the stellar mass function for all snapshots. Four examples of the mass functions and their fittings are presented in \Fig{FigMF}.
As \Fig{FigMF} shows, the fittings are in good agreement with the simulation data above $10^{7}h^{-1}\Msun$.
Moreover, we make successful fittings to all snapshots at $z<10$.
We checked the fitting goodness $R^2$ of each fitting.
In most snapshots, fittings to mass function obtain $R^2$ values above $0.99$.
Eight of them have $R^2$ between $0.95$ and $0.99$.
Four fittings at the earliest redshift have the lowest $R^2$ values, which are between $0.7$ and $0.9$.

\Eqn{EquMF} is built to be a universal function for all redshifts by introducing time $t$ as a hidden variable of the parameters of the original Schechter function.
The time dependence of five parameters ($M_r$, $\Theta_1$, $\Theta_2$, $\alpha_1$ and $\alpha_2$) of \Eqn{EquMF} is presented in \Fig{FigParaMF}.
Their fitting functions  are as follows:
\begin{equation}
	\begin{split}
		&M_s=
		\begin{cases}
			0.0764(t_L-9.74)^2+9.81             & {t_L < 9.74}    \\
			9.81 \ \ \ \ [10^{10}h^{-1}M_\odot] & {t_L \geq 9.74}
		\end{cases} \\
		&log(\Theta_1)=-0.0518e^{0.298t_L}-0.577\\
		&log(\Theta_2)= -0.0749(t_L-7.54)^2-2.94\\
		&\alpha_1=-0.00691(t_L-6.56)^2-0.0207\\
		&\alpha_2=-0.0144(t_L-7.75)^2-0.845
	\end{split}
	\label{EquParaMF}
\end{equation}
Similar to \Eqn{EquFitFr}, we use the looking back time $t_L$ as variable here for convenience.

With the evolved stellar mass function $\Phi_m(m,t)$, we can make a simple approximation to the mean mass growth of galaxies.
For a group of galaxies with the same stellar mass $M$ at time $T$, their frequency density in the whole galaxy population is $\Phi_m(M,T)$.
We ideally consider them to grow at the same pace. After a short time $\Delta t$, each of them gains mass $\Delta m$.
Since $\Delta t$ is such a short time that new galaxies born within this period contribute little to the total population distribution,
the group of galaxies in consideration maintains a constant frequency density at time $T+\Delta t$, which means $\Phi_m(M,T)=\Phi_m(M+\Delta m, T+\Delta t)$.
Then, $\partial m/\partial t$ can be approximated by $\Delta m/\Delta t$.
In practice, we set $\Delta t$ to  $0.001 Gyr$, and determine  $\Delta m$ by finding the pairs of points sharing the same value of $\Phi_m$. \cite{Muzzin2013} used the same method to estimate the average mass growth of individual galaxies.

This approximation assumes a very ideal situation that can hardly be applied to a single galaxy.
However we should note that $\partial m/\partial t$ in \Eqn{EquRM} denotes an average mass change rate.
Regardless of whether individual galaxies are growing via steady accretion or sudden mergers, all these scenarios are taken into account and averaged.
The evolution of the mass function is also the result of average changing trends.
Therefore, it is reasonable to use this approximation.
On the other hand, although it is possible to extract $\partial m/\partial t$ for each galaxy via their mass history  and then obtain the average value in simulations, this method can not be applied to observational data.
We would like to make our method more applicable to observational data.

\Fig{FigRqM} shows our prediction of the mass quenching rate $\lambda_m$.
The pattern of $\lambda_m$ is quite similar to the intrinsic mass quenching rate $\Re_{q,im}$, but has a smaller amplitude.
The maximum value of $\lambda_m$ is  $0.027\ Gyr^{-1}$ at $t_L=3 Gyr$,
and rises to $0.048\ Gyr^{-1}$ at $t_L=0.001 Gyr$.
However, the shape of the peak of $\lambda_m$ is more extended than that of $\Re_{q,im}$, especially at high redshifts.
Their peak positions are also different.
$\lambda_m$ peaks at around $10^{10}$ to $10^{10.6}h^{-1}\Msun$, which is lower than the peak position of $\Re_{q,im}$.
Therefore, compared with intrinsic mass quenching, mass quenching affects  galaxies with relatively lower stellar masses and covers a wider range of galaxies.

\subsection{Environment Quenching Rate}
\label{SsecPQR}

By inserting  \Eqn{EquFitFr} into \Eqn{EquRq}, we obtained the expanded formula of $\lambda_\rho$ as
\begin{equation}
	\begin{aligned}
		\lambda_\rho = & \frac{\partial f_q}{\partial \rho}\frac{\partial \rho}{\partial t}                                      \\
		=              & (1-f_q)\frac{\alpha_\rho}{\rho_c}(\frac{\rho}{\rho_c})^{\alpha_\rho-1} \frac{\partial \rho}{\partial t} \\
		=              & (1-f_q(m,\rho,t_L))\frac{0.546\times10^{-0.0245t_L}}{\rho}                                              \\
		               & \times(1.35\times10^{-0.211t_L-3}\rho)^{0.546\times10^{-0.0245t_L}}                                     \\
		               & \times \frac{\partial \rho}{\partial t}
	\end{aligned}
	\label{EquRP}
\end{equation}

\begin{figure}[htbp]
	\centering
	\includegraphics[width=1\linewidth]{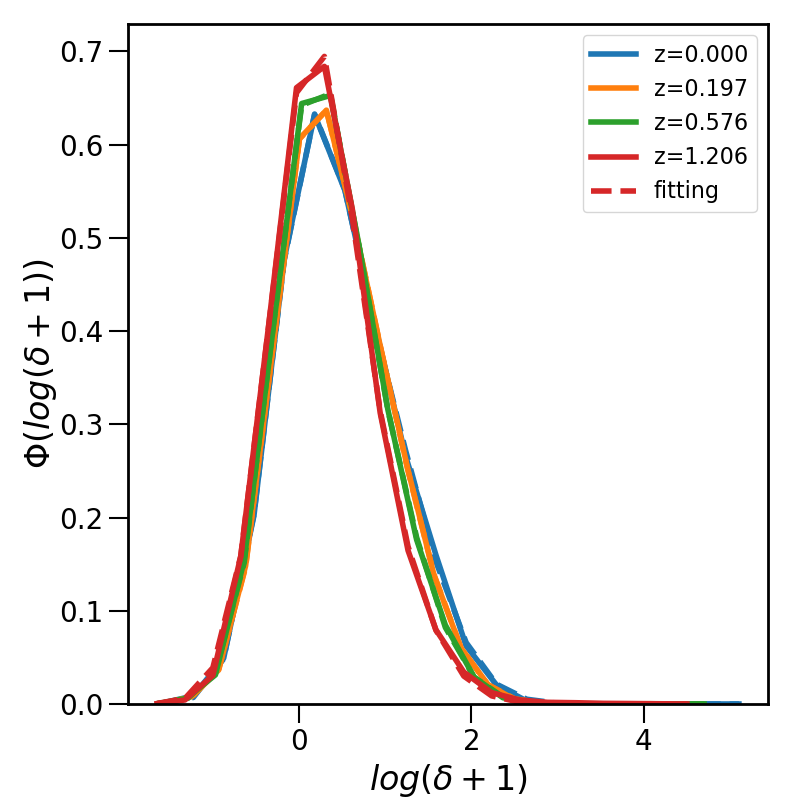}
	\caption{Overdensity function of galaxies in Illustris-1 at different redshifts. Dashed lines with the same color are fitting curves at the same redshifts. }
	\label{FigDF}
\end{figure}

Obviously, the key to evaluating $\lambda_\rho$ is calculating the mean density change rate $\partial \rho/\partial t$.
We use the same method as for the evaluation of $\lambda_m$ in \Ssec{SsecMQR}.
First an evolved overdensity function $\Phi_{\rho}(\rho,t)$ is established, and then $\partial \rho/\partial t$  derived from the change in the overdensity function.
We find a good fitting to the overdensity functions  with two components log-normal function at redshift $z<10$:
\begin{equation}
	\label{EquParaDF}
	\begin{split}
		\Phi_\rho(log\rho,t)=	&\frac{A_1(t)}{\sqrt{2\pi}\sigma_1(t)}e^{-\frac{(log\rho-\mu_1(t))^2}{2\sigma_1(t)^2}} + \\
		&\frac{A_2(t)}{\sqrt{2\pi}\sigma_2(t)}e^{-\frac{(log\rho-\mu_2(t))^2}{2\sigma_2(t)^2}}
	\end{split}
\end{equation}
The overdensity functions and their fittings at five redshifts are shown in \Fig{FigDF} as examples. Most of the fittings have $R^2$ values above $0.98$. The $R^2$ of the worst fitting is $0.73$.

\begin{figure}[htbp]
	\centering
	\includegraphics[width=1\linewidth]{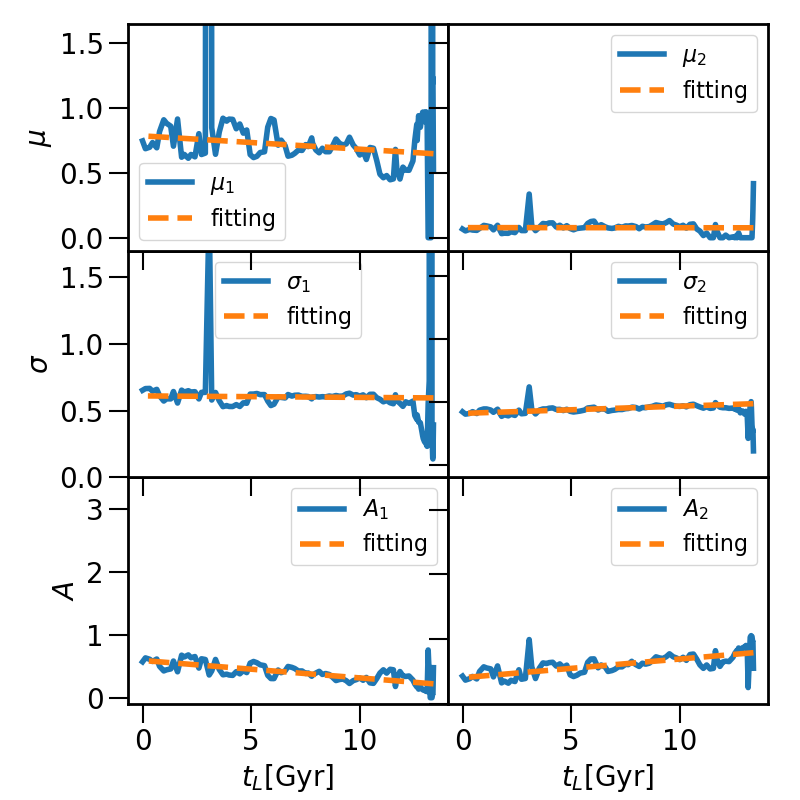}
	\caption{The parameters of the galaxy overdensity function as functions of looking back time $t_L$. The fitting lines of their evolving curves are presented as orange dashed lines.}
	\label{FigParaDF}
\end{figure}

\begin{figure*}[htbp]
	\centering
	\includegraphics[width=1\linewidth]{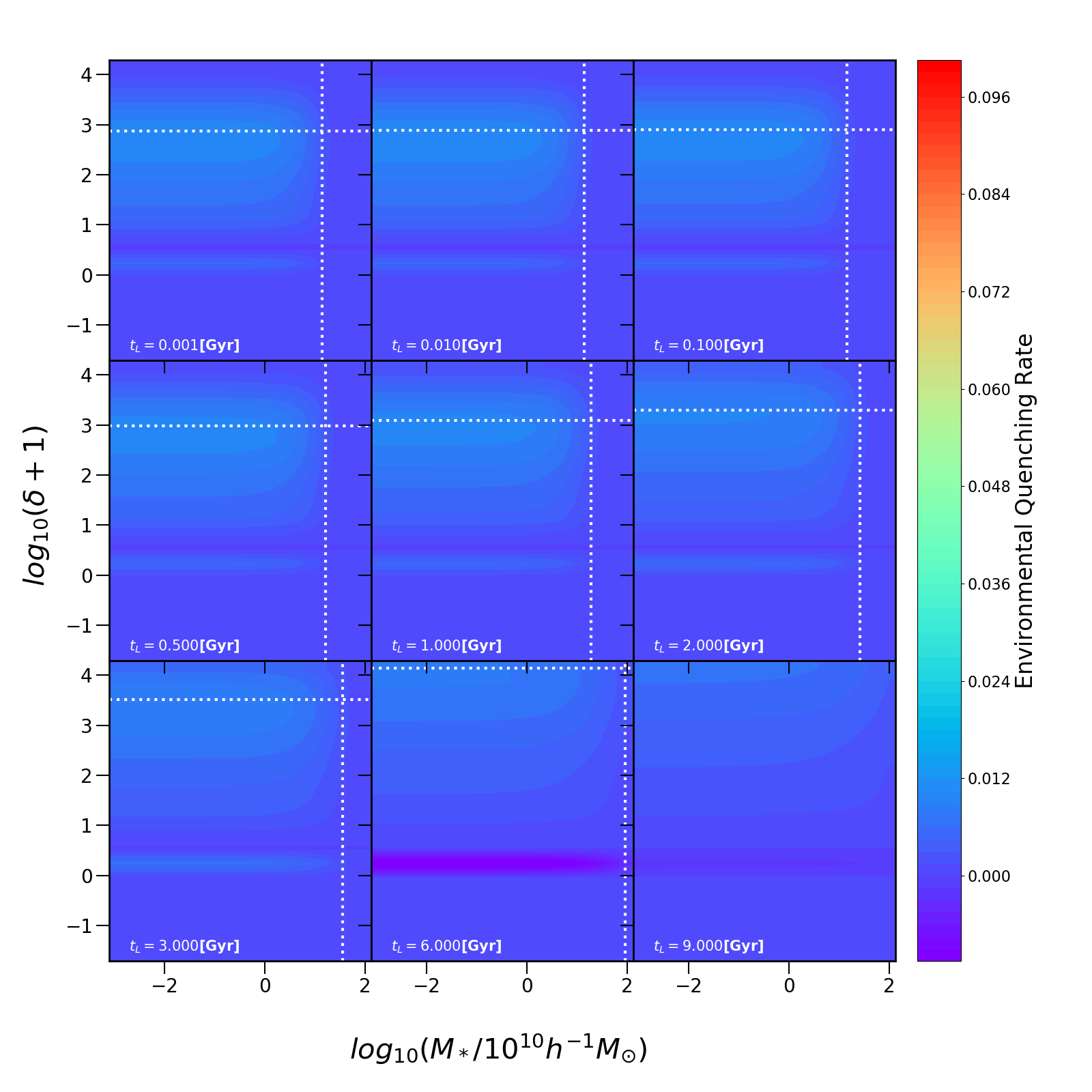}
	\caption{The environmental quenching rate $\lambda_\rho$ as a function of the stellar mass and environmental overdensity of galaxies at different times. The configuration is the same as in \Fig{FigRqIm}.
		The white dotted lines shows  $m_c(t)$ and $\rho_c(t)$ at that time.
	}
	\label{FigRqD}
\end{figure*}

Parameters in \Eqn{EquParaDF} are also regarded as time dependent to introduce evolution to overdensity function.
The fitting results of these parameters as time dependent functions are:
\begin{equation}
	\begin{split}
		&A_1=-0.028t_L+0.59\\
		&\sigma_1=-0.0011t_L+0.61\\
		&\mu_1=-0.010t_L+0.79\\
		&A_2=0.029t_L+0.41 \\
		&\sigma_2=0.0057t_L+0.41\\
		&\mu_2=0.078
	\end{split}
\end{equation}
\Fig{FigParaDF} shows the evolution and fitting curves to these parameters.

Finally we obtain the map of the environmental quenching rate in \Fig{FigRqD}.
Similar to $\Re_{q,i\rho}$, $\lambda_\rho$ is independent of the stellar mass and changes with overdensity, except for the $0$ quenching rate at the high mass end.
In contrast to  other parts of the quenching rate, the environmental quenching rate exhibits two peaks.
One is a narrow peak at $log_{10}(\delta+1)\approx 0.2$.
Another peak center at $log_{10}(\delta+1)\approx 3.3$ emerges when $t_L=3Gyr$
and slowly moves to lower density region with time.
Additionally, the environmental quenching rate is much weaker compared with the other three parts.
Its maximum value is approximately $0.008\ Gyr^{-1}$ and does not change much with time.

\section{Discussion}
\label{sec:compare}
\subsection{Evolution Trend of the Quenching Rate}
From \Sec{sec:quenchz}, we can find several features of the quenching rate in Illustris-1.
The major feature is that it peaks within a specific range.
The peaks of the quenching rate indicate that the quenching process is most efficient there.
Our results show that in Illustris-1,  galaxies with a stellar mass of $10^{10}\sim10^{11}h^{-1}M_\odot$ and an environmental overdensity of $10^{2.8}\sim10^{3.5}$  are much more likely to be quenched.
The main force of the quenching activity moves slowly from high-mass galaxies to low-mass galaxies, from the high-overdensity region to the low-overdensity region.
The downsizing of quenched galaxies has also been reported in many other works \cite[e.g.][]{DeLucia2007,Kodama2007,Rudnick2009}. They found that the passive galaxy population  extended towards lower stellar masses, which is in agreement with our analysis of the Illustris-1 data.

To view the time evolution more clearly and make it comparable with other works, it is worthy examining on the overall
quenching rate at each time.
Here, we make integrate the four parts of \Eqn{EquFr} to obtain the   quenching rate over all galaxies in each snapshot.
\begin{equation}
	\begin{split}
		&RQ_{im}=\frac{\int\Phi_{\rho}\int\Phi_m\Re_{q,im}dmd\rho}{\int\Phi_{\rho}\int\Phi_mdmd\rho} \\
	\end{split}
\end{equation}
\begin{equation}
	\begin{split}
		&RQ_{i\rho}=\frac{\int\Phi_{\rho}\int\Phi_m\Re_{q,i\rho}dmd\rho}{\int\Phi_{\rho}\int\Phi_mdmd\rho} \\
	\end{split}
\end{equation}
\begin{equation}
	\begin{split}
		&RQ_{m}=\frac{\int\Phi_{\rho}\int\Phi_m\lambda_m dmd\rho}{\int\Phi_{\rho}\int\Phi_mdmd\rho} \\
	\end{split}
\end{equation}
\begin{equation}
	\begin{split}
		&RQ_{\rho}=\frac{\int\Phi_{\rho}\int\Phi_m\lambda_\rho dmd\rho}{\int\Phi_{\rho}\int\Phi_mdmd\rho}
	\end{split}
\end{equation}

\begin{figure}[htbp]
	\centering
	\includegraphics[width=1\linewidth]{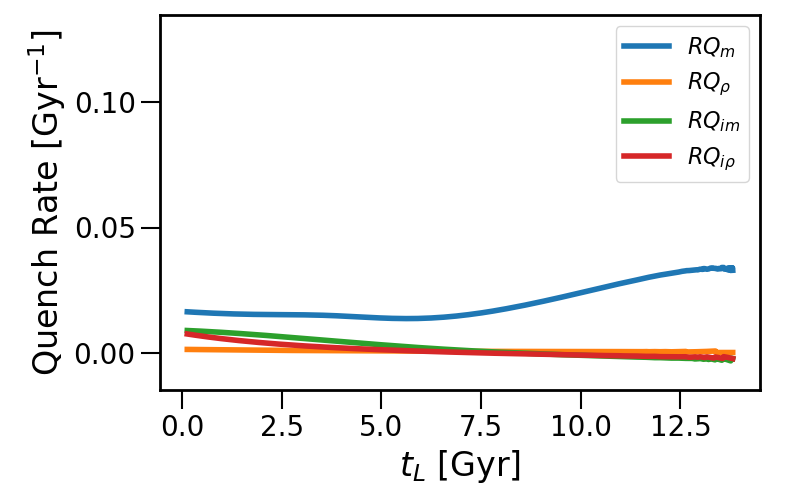}
	\caption{The theoretical overall quenching rate as a function of looking back time $t_L$ predicted by our method. The quenching rates are separated into four parts: intrinsic mass quenching ($RQ_{im}$), intrinsic environmental quenching ($RQ_{i\rho}$), mass quenching ($RQ_m$) and environmental quenching ($RQ_\rho$). Different parts are distinguished by different colors.
	}
	\label{FigNC}
\end{figure}

The overall quenching rate as a functions of looking back time $t_L$ is shown in  \Fig{FigNC}.

From \Fig{FigNC}, we can see that the mass quenching rate dominates the quenching process at all times.
Although \Fig{FigRqIm} and \Fig{FigRqM} suggest that intrinsic mass quenching has
larger peak value than mass quenching, mass quenching takes effect on galaxies
with a much wider stellar mass range than that of intrinsic mass quenching,
leading to the domination of mass quenching when the whole population is  considered.
This outcome is in agreement with \cite{Peng2010}.
According to the curve of $RQ_m$, about $3.34\%$ of galaxies were quenched per $Gyr$ due to mass quenching (including merger quenching)  $12\ Gyr$ ago.
From $t_L\simeq12\ Gry$ to $t_L\simeq6\ Gyr$, the mass quenching rate decreased to $1.36\%\ Gyr^{-1}$, then increased slowly and steadily  to $1.63\%\ Gyr^{-1}$ at present.
The turning  point of this curve,  $t_L\simeq6 Gry$, corresponds to redshift   $z\simeq0.65$ .
\cite{PintosCastro2019} and \cite{Kawinwanichakij2017} showed increasing mass quenching efficiency with decreasing redshift.
The slope of the quenching efficiency $\varepsilon_m$ becomes shallower with decreasing  redshift.
In addition, there is a turning point for $\varepsilon_m(t)$ at $z\sim0.6$.
Considering that $d\varepsilon_m/dt$ is close to $RQ_m+RQ_{im}$,
the agreement between the turning point of $\varepsilon_m$   in \cite{PintosCastro2019} and \cite{Kawinwanichakij2017} and our prediction of the mass quenching rate is not coincident.
From the viewpoint of the mathematics, this turning point comes from the non-monotonicity of parameters $\Theta_2$, $\alpha_1$ and $\alpha_2$ in the stellar mass function.
Their evolution curves have turning points at $t_L=7.54$, $t_L=6.56$ and $t_L=7.75$. (see \Fig{FigParaMF} and \Eqn{EquParaMF}).

Intrinsic mass quenching is the second most effective component.
At very early times, it is approximately $-0.3\%\ Gyr^{-1}$.
The negative quenching rate means that more active galaxies are born than galaxies quenched at that time for galaxies without a significant mass change.
Then, the intrinsic mass quenching rate continues to increase.
It becomes $0$ at $t_L=8.13\ Gyr$ ($z\simeq1.07$), and reaches $0.88\%\ Gyr^{-1}$ (the present value).
The intrinsic environmental quenching rate has the same trend as the intrinsic mass quenching rate but with a slightly smaller amplitude.

The environmental quenching rate is very small.
It evolves from $0$ to $0.1\%\ Gyr^{-1}$, which is approximately $1$ dex below the intrinsic quenching part.
Because the overdensity function changes only slightly, the average overdensity change rate $\partial \rho/\partial t$ is very small.
Therefore, our method predicts a very small environmental quenching rate.
This is also physically reasonable.
Although many environmental effects, such as mergers, ram pressure gas stripping, and galaxy encounters, are considered to be related to galaxy quenching,
a galaxy does not synchronize with the environmental effects (or environmental overdensity changes).
\cite{Pallero2019} found that many satellites are pre-quenched before falling into a cluster or post-quenched afterwards.
In this case, we find a galaxy quenched without instantaneous overdensity variation as we expect, which makes this quenching mathematically an intrinsic environmental quenching rather than environmental quenching.
Since the $RQ_{\rho}$ is much smaller than  $RQ_{i\rho}$,  we consider that most environment related quenching processes, such as satellite quenching, contribute to intrinsic environmental quenching.
Thus, environmental physics may have a negligible pre- or post-processing effect on quenching.

\subsection{Prediction Versus Actual Quenching Process in the Simulation}
In the previous section we presented our analytical formula, which can  describe the evolution of the quenched fraction as a function of the galaxy stellar mass, overdensity, and time in Illustris-1.
The introduction of the time evolution provides a direct view of how much  each aspect influences the quenching rate.
However, there is still an important question: how well does this method work?
To answer this question, we compare our prediction  with the actual data of the quenched fraction in the Illustris-1 simulation.

Limited by the sample volume, galaxy number in each stellar mass bin and overdensity bin is not large enough, leading to significant numerical errors for calculating the quenching rate $\Re_q$ in each bin.
So we only compare the overall quenching rates.
\begin{figure}
	\label{FigNC1}
	\includegraphics[width=1\linewidth]{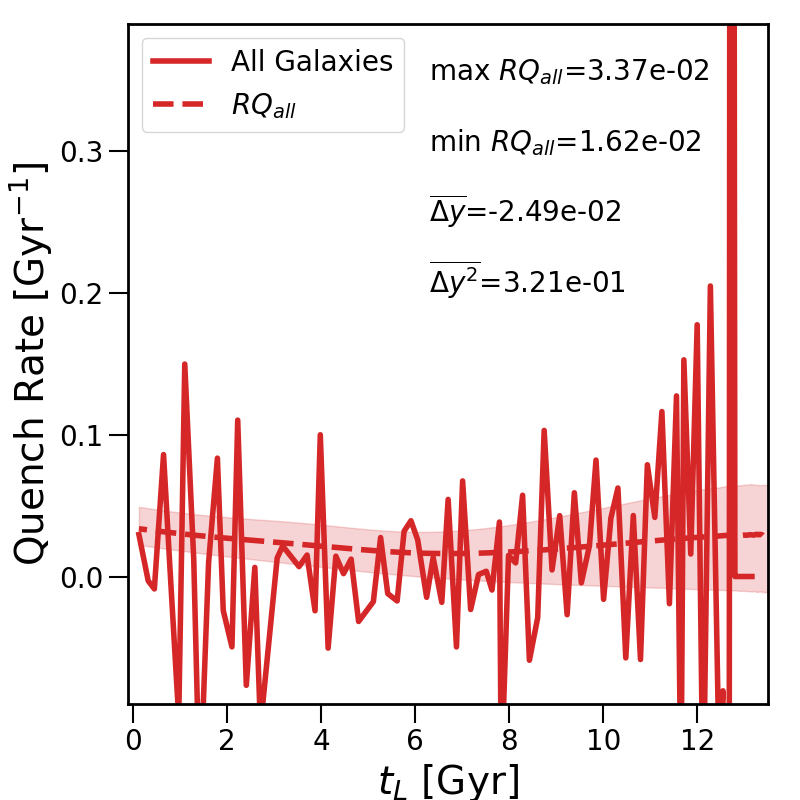}
	\caption{The total quenching rate derived from our prediction (dashed lines) compared with the quenching rate calculated from the history of galaxies in Illustris-1(solid lines).
		The statistic is based on the whole galaxy population.
		The standard error of the theoretical quenching rate is shown as the shaded area in the plot.
		The mean deviation ($\overline{\Delta y}$) and mean root squared deviation($\overline{\Delta y^2}$) between the prediction and actual quenching rate are shown in the figure.
	}
\end{figure}

\begin{figure*}[htbp]
	\centering
	\includegraphics[width=0.3\linewidth]{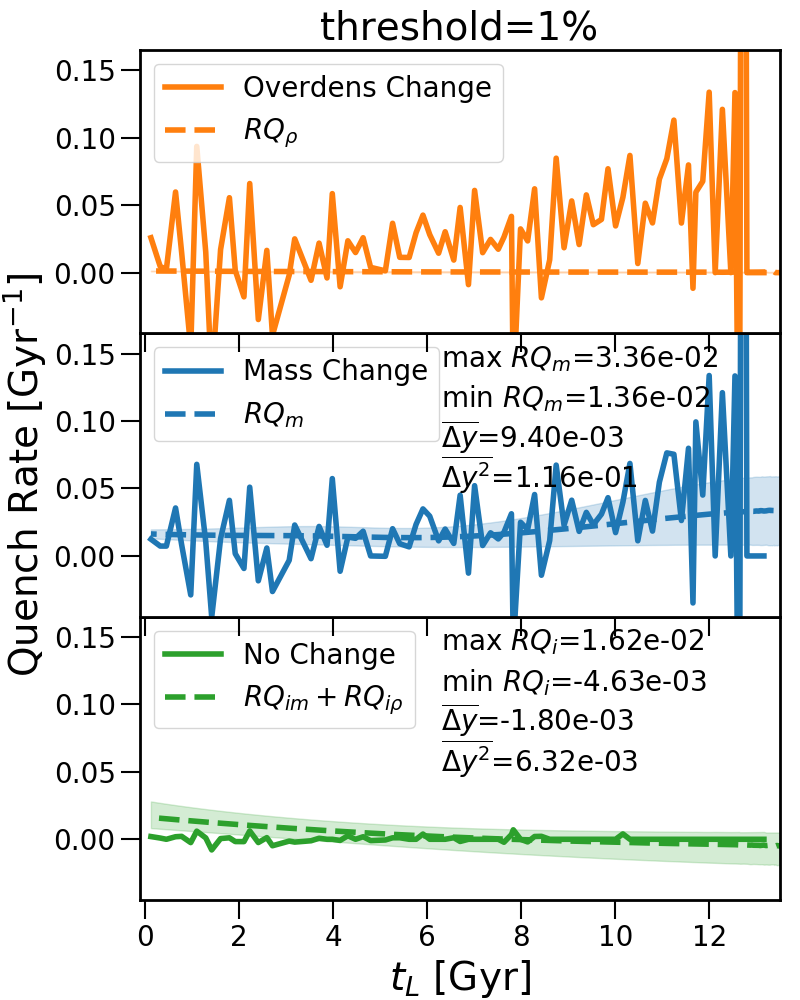}
	\includegraphics[width=0.3\linewidth]{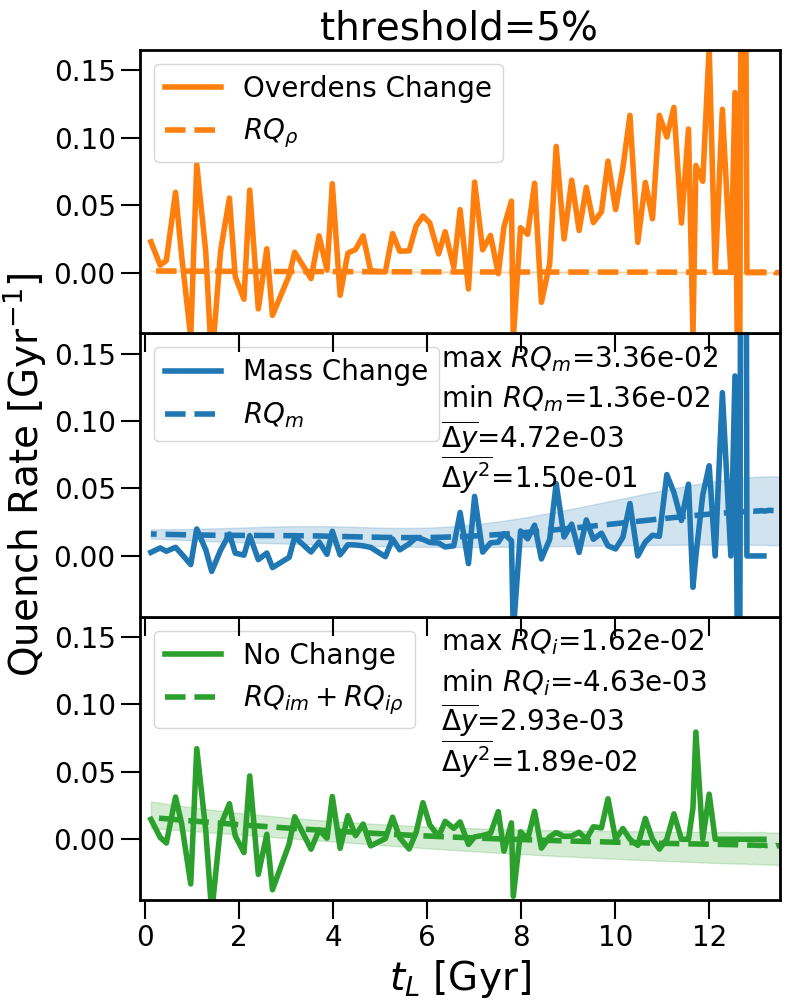}
	\includegraphics[width=0.3\linewidth]{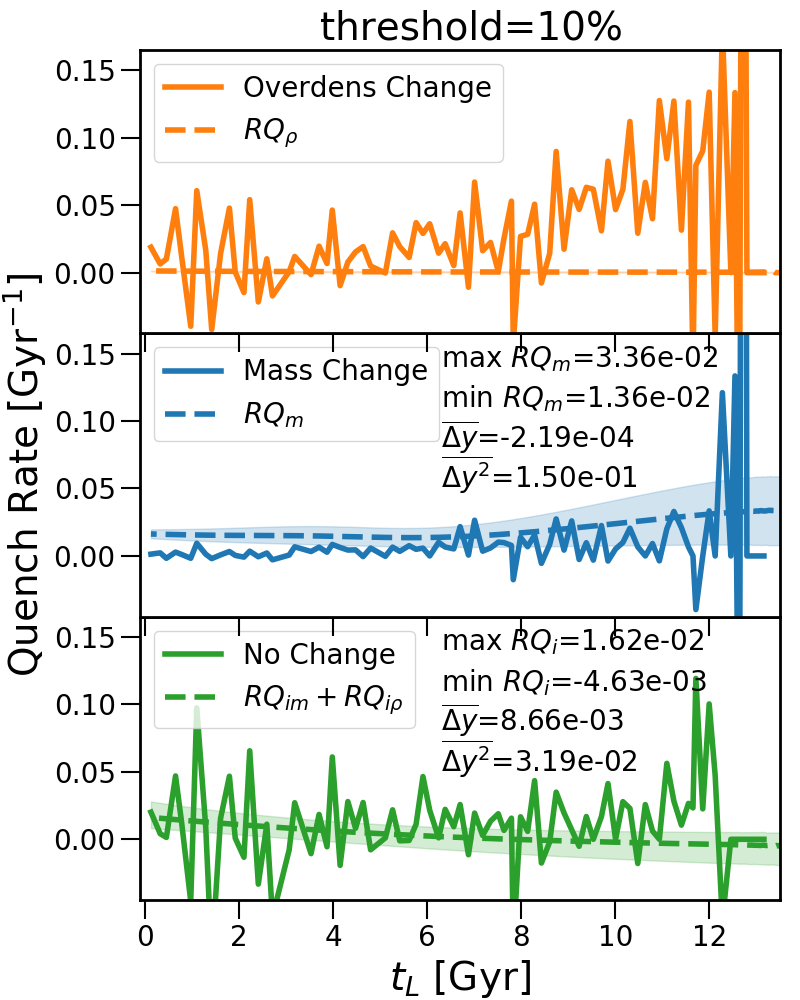}
	\caption{The overall quenching rates of different parts derived from our model (dashed lines) compared with the quenching rate calculated from the history of galaxies in Illustris-1(solid lines).
		The quenching rates for quenching galaxies with environmental overdensity changes , with mass changes  or with no change are plotted from top to bottom .
		The standard error of the theoretical quenching rate is shown as the shaded area in the plots.
		The standard error of theoretical environmental quenching is too small to be visible in the plots.
		The mean deviation ($\overline{\Delta y}$) and root mean squared deviation($\overline{\Delta y^2}$) between the prediction and actual quenching rate are written in the figures.
		For the environmental quenching rate part(top row), the deviation information is omitted since it is very large.
		We use different thresholds to classify whether a galaxy is quenching with a stellar mass change or an environmental overdensity change.
		The results for thresholds of $1\%$, $5\%$ and $10\%$ are shown in the figures from left to right.
	}
	\label{FigNC2}
\end{figure*}

The actual quenching rate can be calculated by counting the number of quenched galaxies in each snapshot with the approximation:
\begin{equation}
	\begin{split}
		\frac{df_q}{dt}&=\frac{d(N_{quench}/N_{total})}{dt} \\
		&=\frac{1}{N_{total}}\frac{\Delta N_{quench}}{\Delta t}-\frac{N_{quench}}{N_{total}^2}\frac{\Delta N_{total}}{\Delta t}
	\end{split}
\end{equation}
$\Delta N_{total}$ is the change in the total number galaxies.
$\Delta N_{quench}$ is the number of newly quenched galaxies minus the number of reactivating galaxies between two successive snapshots.
$\Delta t$ is the time corresponding the time interval in units of Gyr.
By tracing the history of each galaxy, we can distinguish whether the quenching process of a galaxy is accompanied by a stellar mass change or an environmental overdensity change.
With this method, we obtain the actual quenching rate for galaxies with the change in mass, with the change in overdensity  or without any significant changes in mass and overdensity.

First, we check the quenching rate over the whole galaxies population above $10^9M_{\odot} h^{-1}$.
As \Fig{FigNC2} show, the actual quenching rate is very fluctuant. Our prediction roughly goes through the mean value of the actual quenching rate.
The scatter of the actual quenching rate is  far beyond the errors of our prediction.
The error of the theoretical prediction is shown as a shaded area in the figure.
The error is derived from the fitting errors of the parameters of the quenched fraction function and of the fitting functions of the evolution curves of these parameters.
However, we note that the errors for $\partial m/\partial t$ and $\partial \rho/\partial t$ are not taken into account,
as these two parts are derived from the evolved stellar mass function and environmental overdensity function in a numerical way.
With out an analytical formula, it is difficult  to propagate the errors.
Because we count quenching events according to the SFR of a galaxy and its progenitor in neighboring snapshots, the time interval may be too small to encounter too much fake quenching and reviving.
This could enlarge the fluctuation.
On the other hand, a time interval that is too long will eliminate information.
We assume that it would be better if we take reference of the quenching time scale to determine the time interval. Since this improvement is very time consuming, we plan to do so in future works.

We can also check whether our prediction on four quenching modes agrees with the real circumstances.
For actual quenching galaxies, we divide them into three groups,
labeled ``Mass Change'' , ``Overdens Change'' and ``No Change''.
These three groups correspond to $RQ_m$, $RQ_\rho$ and $RQ_{im}+RQ_{i\rho}$, respectively.
For brevity, we label $RQ_{im}+RQ_{i\rho}$ as intrinsic quenching rate in the following context.
In practice, we regard a quenching galaxy with a more than $th\%$ mass change as belonging to the ``Mass Change'' group.
The criteria are the same for the overdensity changes used to build the ``Overdens Change'' group.
If a quenching galaxy has both its mass and environmental overdensity varieties larger than $th\%$, we attribute it half to the ``Mass Change'' group and half to the ``Overdens Change'' group.
If a quenching galaxy has both its mass and environmental overdensity varieties less than $th\%$, it contributes to the ``No Change'' group.
$th$ here is a tunable threshold.
If a galaxy is revived from quenched to star-forming, the corresponding counter minus $1$. Therefore, a negative quenching rate is possible.
Note that, after integration, the quenching rate contributed by mass transformation and overdensity transformation are eliminated from $RQ_m$ and $RQ_\rho$.
Moreover, it is impossible to distinguish the intrinsic mass quenching part and intrinsic environmental quenching part in the simulation.
We test different values of $th$ from $1$ to $10$ and show 3 results in \Fig{FigNC2}.

The actual quenching rates of different modes are shown in  \Fig{FigNC2}
as the solid lines and compared with their theoretical predictions, plotted as dashed lines with the same color.
As \Fig{FigNC2} shows, the prediction of the intrinsic quenching rate ($RQ_{im}+RQ_{i\rho}$)  from our method agrees roughly with the actual data.
However the fluctuation of the actual quenching rate still exceeds the errors in many places.
The mass quenching rate is slightly overestimated at lower $t_L$, but roughly meets the trend at earlier times.
The actual environmental quenching rate fluctuates with a very large amplitude.
Our prediction of the environmental quenching rate is very close to $0$.
It underestimates the environmental quenching rate.
The biases  in the mass quenching rate and environmental quenching rate are due to the average $\partial m/\partial t$ and $\partial \rho/\partial t$ values oversmoothing the changes in mass and environmental overdensity.
The average $\partial m/\partial t$ and $\partial \rho/\partial t$ values derived from the mass function and environmental overdensity function are small because the stellar mass function and environmental overdensity function do not change too much, especially for the environmental overdensity function.
The actual scenario is much more complicated.
For a single galaxy, its stellar mass and environmental overdensity could experience strong variation, while the whole galaxy population maintains its mass function and overdensity function steadily.
Consequently, we find many more galaxies quenching with fluctuated stellar mass or environmental overdensity when we try to trace the history of each galaxy.

\subsection{Link to physical processes}

Although we are trying to derive a quenching rate function in a purely mathematical way, the ultimate goal of establishing this function is to predict the physical drivers of quenching.
Unfortunately, our method can predict only the intrinsic quenching part soundly, which significantly lessens the credibility of exploring the physics with our method.
We plan to explore the physics after refining the method in future work.
In this work, we show some physical evidence supporting our classification on quenching galaxies.

\begin{figure}
	\label{FigCen}
	\includegraphics[width=1\linewidth]{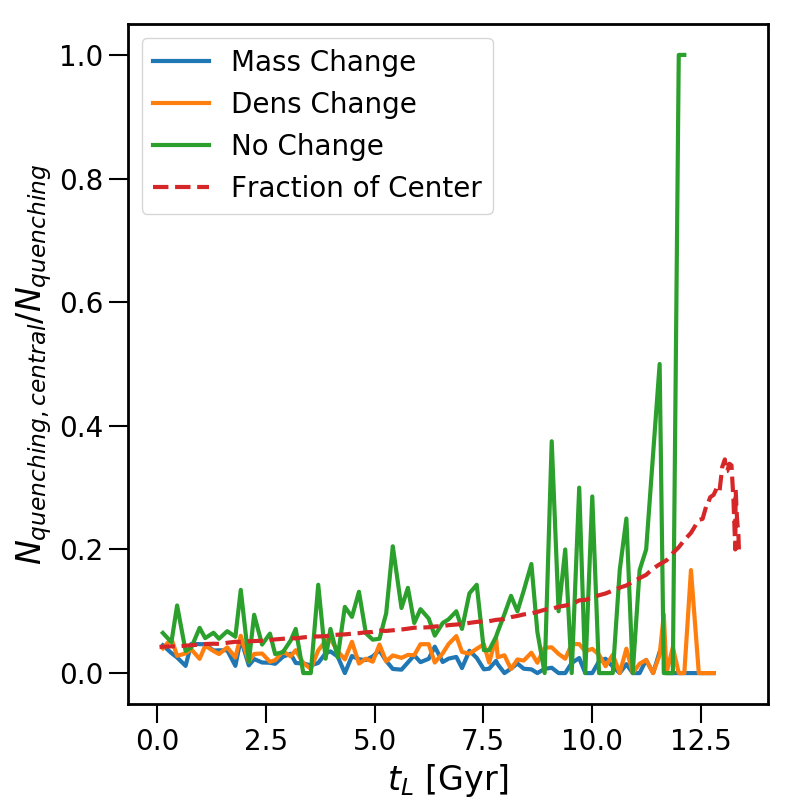}
	\caption{Fraction of central galaxies of newly quenching galaxies at different looking back time. Quenching galaxies with stellar mass changes, environmental overdensity changes or without any changes are shown by lines of different colors. The red dashed line show the fraction of central galaxies of the whole galaxy population at the corresponding time.}
\end{figure}

In our method, there are four different quenching modes: intrinsic mass quenching, intrinsic environmental quenching, mass quenching, and environmental quenching.
In practice, the former two modes can not be distinguished in observations and therefore could be combined into intrinsic quenching.
Theoretically, intrinsic quenching, mass quenching and environmental quenching correspond to galaxy quenching processes without changes in the mass and environments, with changes in mass and with changes in the environments.
We could find quenching galaxies along with their quenching type via merger trees from the Illustris-1 simulation.
By investigating the physics of quenching galaxies at the same time , we find that it is possible that the three quenching modes correspond to different sets of physics.

First, we check the fraction of central galaxies in different quenching modes.
As \Fig{FigCen} shows, very rarely do central galaxies take part in mass quenching(blue line) or environmental quenching(orange line).
Mass quenching and environmental quenching are most likely to take place among satellite galaxies.
In other words, central galaxies prefer the intrinsic quenching mode.
For intrinsic quenching, its fraction of central galaxies is almost the same as the central galaxy fraction of the whole galaxy population (red dashed line).
This implies that satellite galaxies have the same probability of undergoing intrinsic quenching as do central galaxies.

\begin{figure*}
	\label{FigPhy}
	\includegraphics[width=0.33\linewidth]{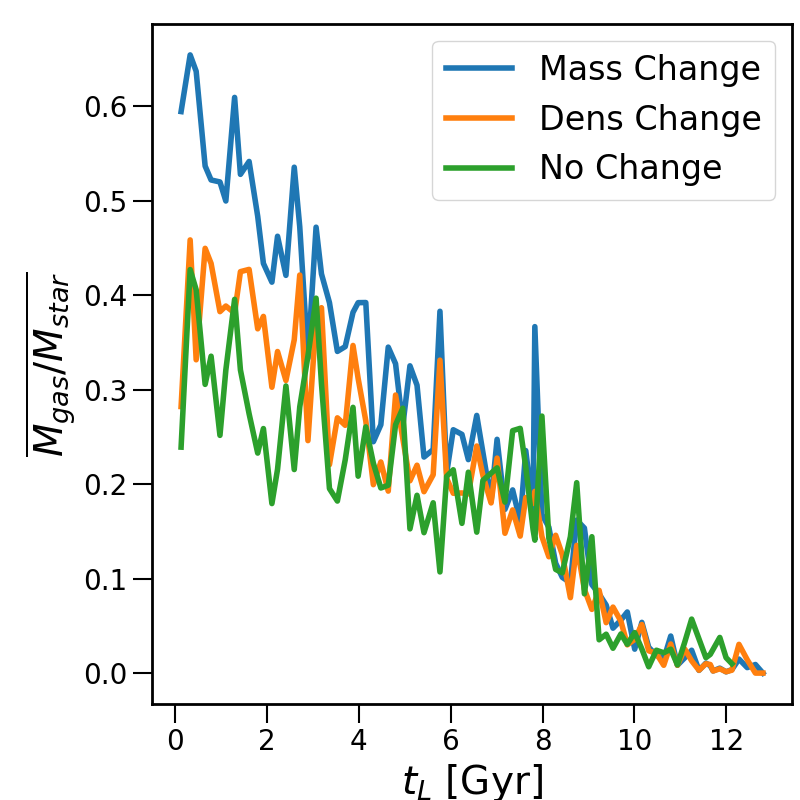}
	\includegraphics[width=0.33\linewidth]{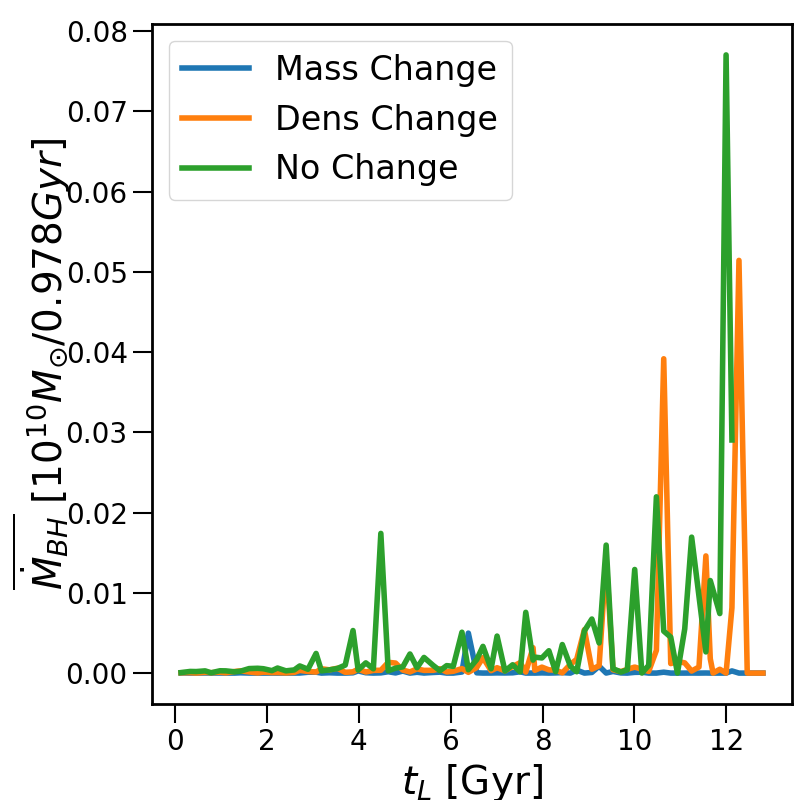}
	\includegraphics[width=0.33\linewidth]{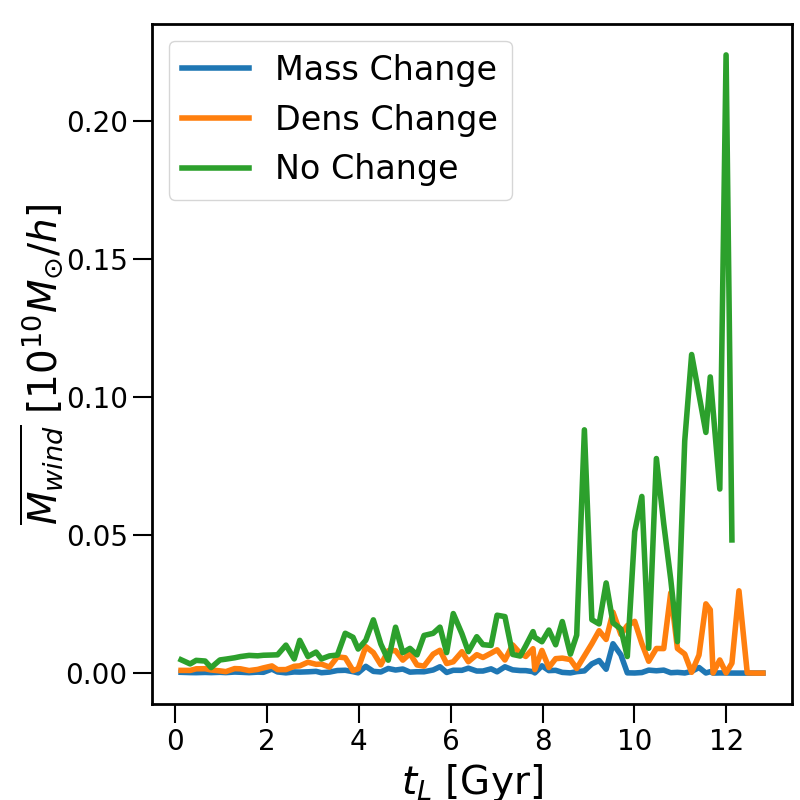}
	\caption{The average gas fraction (left), black hole mass accretion rate (middle) and stellar wind (right) of newly quenching galaxies at different looking back time. Quenching galaxies with stellar mass changes, environmental overdensity changes or without any changes are shown by lines of different colors. }
\end{figure*}

Then, we check the physics of the galaxies in different quenching mode.
We find that the gas fraction, mass accretion rate of black holes and mass of stellar wind are distinguished among different quenching modes.
\Fig{FigPhy} shows the average values of these physical properties of galaxies in different quenching modes.

Mass quenching has almost no black hole accretion and no stellar wind, which means the lack of AGN feedback and stellar feedback.
Mass quenching galaxies are relatively rich in gas.
A reasonable explanation for this scenario is that strong star formation consumes cold gas quickly, resulting in a large stellar mass increment and abundant hot gas remaining.
Intrinsic quenching galaxies have strongest AGN activity and stellar wind, and a least gas fraction.
These features perfectly fit the scenario that a galaxy is quenched due to  gas evaporation by feedbacks.
Environmental quenching is usually considered as a result of gas stripping caused by the surrounding density field or by encountering other galaxies. Thus, it exhibits a relatively smaller gas fraction. On the other hand,  it is probable that inner feedback events occur at the same time with environmental perturbation, which makes the AGN activity and stellar wind visible in this quenching mode.

Here, we show only some rough results. Connecting the physics with the quenching processes requires more work.
Hence, we stop here and will explore this topic in future works.

\subsection{Discussions on Accuracy }
In this part, we briefly discuss the precision of our approach.

First, there is no denying that our approach does not fully achieve our goal.
The expectation is to predict the accurate intrinsic quenching rate, mass quenching rate and environmental quenching rate.
However, our approach recovers only the intrinsic quenching rate.
We assume that this is because the mass change and the environmental change of galaxies have many non-linear evolutions.
In this case, our approximation in \Ssec{SsecMQR} and \Ssec{SsecPQR} is too simplified.
We will try to adjust the mass quenching and environmental quenching parts in future works.
Since these three quenching modes can be separated in terms of both  the mathematics and physics, we can modified the mass quenching and environmental quenching parts separately while keeping the intrinsic quenching part unchanged.

Second, the accuracy of our approach is extremely dependent on how we describe the features of the galaxy population.
We use many functions to fit the distribution of the quenched fraction, stellar mass and environmental overdensity of galaxies and functions to fit the evolution trends of all parameters of the distribution functions above.
All these fittings are the best fittings to the Illustris-1 data in our opinion, but they may not satisfy everyone.
For example, many works claimed that the mass quenching efficiency and environmental quenching efficiency are not separable \citep[e.g.,][]{Contini2019}.
In this case \Eqn{EquFr} should be modified.
One possible modification is to add a cross item of $\rho$ and $m$  to \Eqn{EquFr} :
\begin{equation}
	\label{EquFr2}
	f_{q}(\rho,m)=1-e^{-(\frac{\rho}{\rho_c})^{\alpha_\rho}-(\frac{m}{m_c})^{\alpha_m}-\frac{m^a\rho^b}{c}}
\end{equation}
Alternatively, even more sophisticated formulas could be applied to fit the distribution of the quenched fraction.
\Eqn{EquFr2} is not applied in this work because the fitting program suggests very small factors of the cross item, which means that the cross part of $\rho$ and $m$ could be ignored.
We regard \Eqn{EquFr} as the best quenched fraction function for the Illustris-1 data.
We  emphasize that the format of the fitting functions could be changed freely in the processes of our approach.
The core concept of our approach is determining the quenching rate by $df_q/dt$.
For different physics models, the formats of $f_q$ and its evolution trend could be different.
We should use the appropriate fitting functions according to the practical circumstances.

In this work, the independent variables of quenched fraction function  chosen are the stellar mass of the galaxy and environmental galaxy number overdensity.
They are very traditional variables in exploring galaxy quenching, and easy to measure in observation.
However, this does not mean that the quenched fraction function is limited to these two variables.
In recent years, many researchers have claimed that some other indicators, i.e., bulge mass, black hole mass and $B/T$, are more closer related to the quenched fraction \citep{Wuyts2011, Cheung2012,Teimoorinia2016} .
Many researchers prefer using the centric distance or host dark matter halo mass to describe the environments \citep{Contini2020, Xie2020}.
In this case, the variable of the quenched fraction function could be changed.
Principally, a quenched fraction function with other variables, such as $f_q(M_{bulge}, M_{halo})$, could also work with our approach for calculating the quench rate.
The choice of variables depends on which qualities you want to explore.
If you only want to explore the relation between quenching and black hole mass, then a single variable function $f_q(M_{halo})$ is sufficient.
Information on other relations may be degenerated into the relation between quenched fraction and black hole mass.
In a very extreme case, if you want to explore all the quantities we have mentioned, then
a sophisticated quenched fraction function with multiple variables like $f_q(M_*, M_{bulge}, B/T, M_{BH}, M_{halo}, \delta, r_{centric})$ does work.
However, the readers should remember that more variables do not essentially provide more information.
If two variables are highly related, they can not be used to construct an orthogonal coordinate system.
In this case, using either one to be the variable is equivalent to using the other.
It is redundant to use both of them as the variables.
For example, in the Illustris-1 simulation, the stellar mass, bulge mass and black hole mass of a galaxy are highly related.
Figure 5 of \cite{Sijacki2015} shows a significant linear relation between the bulge mass and black hole mass.
\cite{Sijacki2015} adopted the total stellar mass within the half-mass radius as a proxy for the bulge mass, which is also linearly related to the stellar mass of galaxies we used in this work.
According the the chain rule of differential, $\partial f/\partial M_* = \partial f/\partial M_{bulge}\cdot\partial M_{bulge}/\partial M_*= \partial f/\partial M_{BH}\cdot\partial M_{BH}/\partial M_*$.
The differences of changing variable is only a factor $\partial M_{bulge}/\partial M_*$ or $\partial M_{BH}/\partial M_*$.
When $M_*$, $M_{bulge}$ and $M_{BH}$ are linearly related, this factor is a constant.
Therefore, changing the variable $M_*$ to $M_{bulge}$ or $M_{BH}$ will offer little change in this work.
However, when our approach is applied to other data that do not have such strong linear relations, it might be worth adding $M_{bulge}$ and $M_{BH}$ to the list of variables of quenched fraction function.

\section{Conclusion}
\label{sec:con}
In this work, we explore a method to derive the quenching rate, i.e., the change in the red fraction of galaxies per unit time, of galaxies in the Illustris-1 simulation. By exploring the quenching rate, we can have an intuitive sense of how fast a population of galaxies is quenched.

Our method essentially involves two steps:
\begin{itemize}
	\item [1)]
	      Building up a universal time dependent quenched fraction function $f_q(m,\rho,t)$ according the quenched fraction of galaxies at different times.
	\item [2)]
	      Calculate the quenching rate by calculating $df_q/dt$
\end{itemize}
According to the term of $df_q/dt$, we separate the quenching rate into four parts:
intrinsic mass quenching, intrinsic environmental quenching, mass quenching and environmental quenching. These four parts show very featured patterns:
\begin{itemize}
	\item
	      Intrinsic mass quenching occurs most strongly in galaxies with a stellar mass range of $10^{10.9}\ \sim\ 10^{11.4} M_{\odot}h^{-1} $ at present. The peak value of intrinsic mass quenching is $0.13\ Gyr^{-1}$.
	      The intrinsic mass quenching galaxies shift to a higher stellar mass at an earlier time.
	\item
	      Intrinsic environmental quenching occurs most strongly in  galaxies with an environmental overdensity range of $10^{2.3}\ \sim\ 10^{3.4}$ at present. The peak value of intrinsic environmental quenching is $0.12\ Gyr^{-1}$. The intrinsic environmental quenching galaxies shift to higher environmental overdensity at earlier times.
	\item
	      Mass quenching occurs most strongly in  galaxies with stellar mass range of $10^{9.5}\ \sim \ 10^{11.5}	M_{\odot}h^{-1} $. The peak value is $0.048\ Gyr^{-1}$. At earlier times, the stellar mass range of mass quenching galaxies becomes slightly broader, while the center of the section remains stable.
	\item
	      Environmental quenching occurs in a wide environmental overdensity range from $10^{0.1}$ to $10^{3.5}$.
	      The environmental quenching has two peaks at $log_{10}(\delta+1)\approx 2.7$ and  at $log_{10}(\delta+1)\approx 0.2$.
	      The largest value of environmental quenching rate at present is $0.008\ Gyr^{-1}$.
	      The peak in the high density region shifts to a higher density region at earlier time.
	      The peak of $log_{10}(\delta+1)\approx 0.2$ stays the same position from $t_L=0 Gyr$ to $t_L=3 Gyr$ and becomes almost invisible at earlier times.
\end{itemize}

Then, we perform a simple analysis of the quenching history of the whole galaxy population in the Illustris-1 simulation.
We find that mass quenching in Illustris-1 dominates the quenching process , which is in agreement with some observations \citep[e.g.,][]{Peng2010} but inconsistent with others \citep[e.g.,][]{Balogh2016}.
Intrinsic mass and environmental quenching are the second and third most effective components.
Environmental quenching is very weak.

To validate our method, we compared our prediction with actual quenching rate calculated by counting the number of quenching galaxies at each snapshot.
The actual quenching rate is very fluctuant, but our prediction roughly agrees with the mean value.
Predictions on different quenching modes are also compared with corresponding groups of quenching galaxies.
Our method predicts the actual intrinsic quenching rate well, slightly over-estimates actual mass quenching rate and highly underestimates the actual environmental quenching rate.
We assume that the bias mainly comes from estimating the mean mass change rate and overdensity change rate in our approach.
We will test and improve it in future works.

The mechanism of quenching is largely discussed but still controversial.
It is important to clarify which physical mechanisms are responsible for quenching and how much work they do at different times.
We consider the quenching rate to be an indicator more directly related to instantaneous physical activities;
thus, we tried to find an analytical term of the quenching rate and expect this method to provide a new way to explore the mechanisms of galaxy quenching.
Currently this method is still in a very early state.
To improve it, we have to compensate for the bias in predicting mass quenching rate and environmental quenching rate, to test this method in more simulations with more physical models, and to explore the links between the mathematical terms and physics.

\normalem
\begin{acknowledgements}
	\section*{Acknowledgements}

	The authors thank the Illustris projects for providing the data.
	The authors thank the referee for the constructive comments and suggestions.

	Y.W. is supported by NSFC grant No.11803095 and NSFC grant No.11733010.

	W.P.L acknowledge support from the National Key Program for Science and Technology Research and Development (2017YFB0203300), the National Key Basic Research Program of China (No. 2015C857001) and the NSFC grant (No.12073089).

	W.S.Z. is supported by  NSFC grant 11673077

	L.T is also supported by the Natural Science Foundation of China (No. 12003079) and the Fundamental Research Funds for the Central Universities, Sun Yat-sen University (71000- 31610036)

	Most of the calculations of this work were performed on the Kunlun HPC in SPA, SYSU.

\end{acknowledgements}

\bibliographystyle{aasjournal}
\bibliography{Lib}
\label{lastpage}
\end{document}